**TPRC Conference DRAFT - Subject to revision**

# ICANN and Antitrust[1]

A. Michael Froomkin[2] & Mark A. Lemley[3]

In the late 1990s, the U.S. government ceded de facto technical and policy control over the Internet domain name system (DNS) to a private non-profit company. That company, the Internet Corporation for Assigned Names and Numbers (ICANN), operates under various contracts to the US government, and via additional contracts it has signed with registries and registrars. ICANN acts as the de facto regulator for DNS policy. In addition to technical policy coordination that prevents multiple registries from attempting to deploy colliding top-level-domains (TLDs),[4] ICANN also engages in policies that strongly resemble traditional regulation of market structure. It decides what TLDs will be made available to users, and which registrars will be permitted to offer those TLDs for sale. And it makes those decisions at least in part based on the potential registrars' willingness to offer a package of services that includes mandatory trademark arbitration.

ICANN's curious status as a quasi-non governmental organization with strong ties to the government has occasioned a good deal of comment, much of it negative.[5] Some of that comment has focused on ICANN's rather fitful starts towards open and participatory governance of the DNS.[6] Others have questioned the legitimacy of handing important policy questions over to a private -- or at least mostly private -- entity.[7] In this article, we focus on a hitherto-neglected implication of ICANN's private status: its potential liability under the U.S. antitrust laws, and the liability of those who transact with it.

---

[1] © 2001 A. Michael Froomkin & Mark A. Lemley. All rights reserved.

[2] Professor of Law, University of Miami School of Law. I should disclose that I was involved in some of the events discussed in this paper.

[3] Professor of Law, Boalt Hall, University of California at Berkeley; of counsel, Fish & Richardson PC. We would like to thank Collen Chien for research assistance.

[4] Top-level domains include international domains such as .com and .net and country-code domains such as .uk and .ca.

[5] For a detailed discussion of those ties, and the history of ICANN more generally, *see* A. Michael Froomkin, *Wrong Turn in Cyberspace: Using ICANN to Route Around the APA and the Constitution*, 50 **Duke L.J.** 17 (2000); Jonathan Weinberg, ICANN and the Problem of Legitimacy, 50 DUKE L.J. 187 (2000), *available online* http://www.law.duke.edu/shell/cite.pl?50+Duke+L.+J.+187 . *See also* Graeme Dinwoodie & Lawrence Helfer, *ICANN*, __ **Wm. & Mary L. Rev.** __ (forthcoming 2001); Jay P. Kesan & Rajiv C. Shah, *Fool Us Once Shame on You – Fool Us Twice Shame on Us: What We Can Learn From the Privatizations of the Internet Backbone Network and the Domain Name System*, 79 **Wash. U.L.Q.** 89 (2001); Jessica Litman, *The DNS Wars: Trademarks and the Internet Domain Name System*, 4 **J. Sm. & Emerging Bus. L.** 149 (2000); Timothy Wu, *Application-Centered Internet Analysis*, 85 **Va. L. Rev.** 1163 (1999).

[6] *E.g.* Weinberg, *supra* note 5; Jonathan Weinberg, *Greeks and Geeks*, http://www.law.wayne.edu/weinberg/geeksandgreeks.pdf

ICANN finds itself on the horns of a dilemma. It has some of the indicia of a government corporation exercising authority granted to it by the Department of Commerce.[8] If ICANN is therefore a state entity, it is subject to both constitutional and statutory constraints on its regulatory authority -- notably the requirement of due process and the burdens of the Administrative Procedures Act.[9] ICANN's actions to date are unlikely to satisfy those procedural constraints. Both ICANN and the U.S. government argue that ICANN is not subject to those rules because it is a private industry self-regulatory body.[10] Although one of us has argued that ICANN is best understood to be a state actor for constitutional law purposes, the contrary argument that ICANN is private is not without merit. If ICANN is private, however, it follows that both ICANN and private actors who have relationships with it, are subject to U.S. (and probably non-US) antitrust law.

Previous legal efforts to subject IANA and Network Solutions (NSI) to U.S. antitrust liability have uniformly failed, in part because the courts concluded that the pre-ICANN DNS was run by state actors or those (such as NSI) acting at their behest and was therefore immune from antitrust scrutiny.[11] ICANN argues, however, that it is not as closely tied to the government as NSI was in the days before ICANN was created. If this is correct, it seems likely that ICANN will not benefit from the same immunity. Rather, it will have to defend its actions on their competitive merits. In this article we consider the antitrust implications of ICANN's actions, both for it and for those who interact with it.

Some of ICANN's regulatory actions, although perhaps justified on other grounds, clearly raise competitive concerns. For example, in the recent round of applications for new TLDs, ICANN made it a prime requirement that applicants demonstrate that their proposals would not enable competitive (alternate) roots. Similarly, ICANN prevents certain types of non-price competition among registrars by requiring that they adhere to an ICANN dispute policy (the UDRP) that was, in substantial part, drafted by a lawyer hired by a consortium of existing registrars. ICANN's rules -- coupled with its de facto control over the DNS -- may have the effect of restraining competition.

Further, the process by which those rules were adopted might be characterized as anticompetitive collusion by existing registrars. Those registrars will likely not be subject to the *Noerr-Pennington* lobbying exemption that would shield them from antitrust immunity were ICANN a public body.[12] ICANN sets these and other policies within a structure that gives the 'regulated' a strong voice in the policies applied to them: half of the ICANN Board seats are allocated under corporatist principles by representatives of industries and groups potentially affected by ICANN's actions.

---

[7] *See* Froomkin, *supra* note 5, at 141-59 (questioning the legality of the government's delegation of policy authority to ICANN); Joseph P. Liu, *Legitimacy and Authority in Internet Coordination: A Domain Name Case Study*, 74 **Ind. L.J.** 587, 604 (1999); Paul Schiff Berman, *Cyberspace and the State Action Debate: The Cultural Value of Applying Constitutional Norms to "Private" Regulation*, 71 **U. Colo. L. Rev.** 1263 (2000); Jonathan Zittrain, *ICANN: Between the Public and the Private*, 14 **Berkeley Tech. L.J.** 1071 (1999).

[8] On government corporations see generally A. Michael Froomkin, Reinventing the Government Corporation 1995 **U. Ill. L. Rev.** 543, http://www.law.miami.edu/~froomkin/articles/reinvent.htm.

[9] 5 U.S.C. § 551 et seq.; *see* Froomkin, *supra* note 5 at 94-96 (summarizing caselaw).

[10] *See* Froomkin, *supra* note 5, at 94-96.

[11] *See infra* notes __-__ and accompanying text (discussing these cases).

Whether ICANN has in fact violated the antitrust laws depends on whether it is an antitrust state actor (even if it is not, by hypothesis, a state actor for purposes of constitutional law), whether the DNS is an essential facility, on the extent of ICANN's government contractor immunity under an unusual zero-dollar procurement contract, and on whether it can shelter under precedents that protect standard-setting bodies. Particular attention under antitrust law will be paid to the extent to which the government has both clearly articulated a policy direction for ICANN and actively supervised ICANN's performance of its duties. While Congress is investigating ICANN right now, a plausible argument can be made that the government has neither clearly articulated support for nor actively supervised ICANN's anticompetitive actions. If so, it will be treated just as any other private market participant would be treated.

The exposure of those who take part in ICANN's constituencies, and which lobby ICANN in a collective manner to impose rules on them and their competitors, depends in part on ICANN's status as a quasi-state actor. It also depends on the nature of their conduct, how closely they coordinate behavior, and the effect on competition.

If (as seems likely) a private ICANN and those who petition it are subject to antitrust law, everyone involved in the process needs to review their conduct with an eye towards legal liability. ICANN should act very differently with respect to both the UDRP and the competitive roots if it is to avoid restraining trade. Whether ICANN's UDRP and competitive root policies are desirable is another matter, one on which the authors do not necessarily agree. But if ICANN's policies are in fact desirable, then it seems clear that the government must take a more active role both in setting those policies and in supervising their implementation, either by exercising control over ICANN or by replacing it with a true governmental decision-maker.

In Part I, we briefly review ICANN's history and its relationship to the U.S. government. In Part II, we discuss the antitrust state action doctrine, and how that doctrine has been applied in the past to the DNS. In Part III, we focus on two ICANN policies that are potentially anticompetitive: its restriction on the deployment of new TLDs and its decision to require a uniform arbitration procedure for resolving trademark disputes. Finally, in Part IV we consider the liability of other private parties who petition ICANN to set its policies.

**Part I: ICANN in a Nutshell**

---

[12] Antitrust immunity grew out of *Eastern Railroad Presidents Conference v. Noerr Motor Freight, In*c., 365 U.S. 127 (1961) and *United Mine Workers v. Pennington*, 381 U.S. 657 (1965). For background on the source of the doctrine, see David McGowan & Mark A. Lemley, *Antitrust Immunity: State Action and Federalism, Petitioning and the First Amendment*, 17 **Harv. J. L. & Pub. Pol'y** 293 (1994).

ICANN is a complicated answer to two problems, one technical and one political. The technical problem results from the architecture of the domain name system (DNS) on which the smooth functioning of the Internet relies. The Internet is a giant network of machines that use common protocols to communicate with one another. Every resource on the network has a unique address, called an IP number.[13] Because IP numbers are hard for people to remember, Internet standards provide for the creation of mnemonic names for resources. The DNS is the name given to the complex system for registering those mnemonics -- domain names -- and maintaining the vast distributed, highly cached, directories that permit every browser pointed at a URL to look up the correct IP number and deliver an Internet communication, and every email to reach its destination. The act of looking up a domain name and retrieving the associated IP numbers called *name resolution*.

The original design of the DNS assumed that there would be one hierarchically organized set of domain names, and that every domain name in it would be unique. Unique domain names ensure that every user of the Internet who types a particular URL will find that it resolves to the same IP number associated with that URL, and thus allows a connection to the same resource. The failure to ensure uniqueness -- to allow a condition where different users typing the same thing get routed to different IP numbers -- has been called a "name collision" or, more pejoratively, "instability of the Internet". To avoid these problems, the DNS relies on a system of layered registrations. In what is sometimes called the *legacy DNS* - the DNS that today almost everyone uses -- there is one master file called the *root file* that lists the approved top level domains (TLDs). Each line in this file contains the name of a TLD and the IP number of a computer that has the authoritative *registry* for that TLD. The root file is copied by the thirteen *root servers*, which are the computers that actually resolve any TLD queries that cannot be resolved in hierarchically organized, cached, databases closer to the user.

So long as everyone relies on the same family of hierarchically organized databases, whoever controls the root file enjoys the power to determine which TLDs are accessible to the entire Internet, and what registry's database will be considered the authoritative source of information for that TLD. The database of registrations in each TLD is in turn controlled by a single registry.[14] In contrast, today a group of highly competitive *registrars* provide the service of selling actual entries, or registrations, into the registries. To get the right to use a domain name,[15] a registrant in a gTLD first must find an available name. She then pays a registrar to inscribe her name, contact information, and IP number in the registry. Domain names in .com, .org, and .net are allocated on a first-come, first-serve basis; names in the newer gTLDs will be allocated in more complex fashions that give priority to trademark holders, and also seek to level the playing field for similarly situated applicants competing for a name during the initial "landrush" period when registrations open.[16]

---

[13] Some IP numbers are 'static', i.e. normally unchanging; other are 'dynamic' i.e. shared out and then withdrawn on as-needed bases, e.g. for the length of a dial-up connection via modem. E-mail connections work on a very similar system, although the IP number used to route e-mail is called an "MX" (mail exchange) number. MX numbers, often identical to the IP number, are assigned to domain names (often automatically) via the same process as the ordinary IP number.

[14] There is no technical reason why one registry cannot control multiple gTLDs -- and indeed, VeriSign currently controls three.

[15] Whether a registrant acquires a property interest in a domain name, or merely enjoys a service contract is a controversial question. To date the trend seems to be away from a property right, even though the right is something that can be subject to an in rem action.

Thus, the legacy DNS system has two chokepoints: whoever controls the root controls which, and how many, TLDs will be accessible to the vast majority of Internet users. And, while there can be many competing registrars, and many TLDs competing with each other, each TLD must have one master registry.

The existence of these chokepoints over the legacy root created a political problem. In effect, whoever controlled the root file controlled both whether a given TLD could be part of the Internet and who got the potentially lucrative job of running the TLD's registry. And, by 1997, these were increasingly controversial questions--questions that had landed in the lap of the U.S. government, which found itself controlling the root.[17]

From the viewpoint of high level policy makers, this power was a not-entirely-welcome accident. A series of largely informal arrangements, mostly coordinated by one person, Dr. Jon Postel, and supported by a series of first military, then National Science Foundation contracts,[18] had morphed into something unforeseen, important, and increasingly controversial. DNS management, notably the registration of names in .com, the most popular gTLD, had taken on a commercial life of its own due to an explosion of interest in the Internet. Thanks in part to marketing efforts by Network Solutions Inc. (NSI) -- which enjoyed a government-granted monopoly on selling domain name registrations in the three major gTLDs, .com, .org and .net -- as new users flooded onto the Internet they registered millions of domain names. Meanwhile, however, the very informal, consensus-driven, and perhaps unsophisticated method for creating new gTLDs broke down under the strain of competing interests. Registrars wishing to compete with NSI chafed at its contractual government monopoly, and at its monopolistic practices. Users who were not first to the Internet wanted new, short, domain names, and new gTLDs to register them in if the ones in the existing gTLDs (and especially .com) were already taken. Trademark holders, waking up to the marketing and commercial potential of the Internet, wanted controls on the ability of others to register words that paralleled or resembled their marks. Registrants accused of cybersquatting wanted a less-draconian method of dealing with such charges than NSI's policy of simply de-activating their domain name pending the (slow, sometimes endlessly deferred) resolution of the dispute.

Whether there should be new top-level domain names was especially controversial. Although it is easy for the DNS system controller to create new world-readable gTLDs,[19] and indeed Dr. Postel proposed creating hundreds, intellectual property rights holders objected to additional gTLDs, arguing with some justification that they already faced mounting problems from cybersquatters--speculators who registered domain names corresponding to trademarks and sought to resell them to the trademark holders for profit.[20] Meanwhile, foreign governments, and especially the European Union, began to express understandable concern about the United States' control of a critical element of a global communication and commercial resource on which they foresaw their economies and societies becoming ever-more dependent.[21]

---

[16] *See* http://nameengine.com/dotprotect/timeline.biz.pdf ; http://nameengine.com/dotprotect/timeline.info.pdf

[17] *See* Froomkin, *supra* note 5, at 51-62.

[18] *See generally* Vint Cerf, *A Brief History of the Internet and Related Networks*, Internet Society, *at* http://www.isoc.org/internet-history/cerf.html (July 1995) (documenting the creation and growth of the Internet); Barry M. Leiner et al., *A Brief History of the Internet*, Internet Society, *at* http://www.isoc.org/internet-history/brief.html (reviewing the origin and fundamental ideas behind the Internet).

[19] *See* Froomkin, *supra* note 5, at 22 n. 12.

In June, 1998, a task force headed by Senior Presidential Adviser Ira Magaziner issued the *Statement of Policy on the Privatization of Internet Domain Name System*, known as the DNS White Paper.[22] The White Paper called for the government to transition its control of the DNS to a private corporation identified only as 'NewCo'. The White Paper did not actually mandate the creation of this corporation, but -- nicely skirting the prohibitions of the Government Corporation Control Act[23] -- only said how nice it would be if someone would form it to undertake certain specified tasks so that the government could strike a deal with it.[24] And, not quite fortuitously, a group of worthies did just that, forming a California non-profit corporation called ICANN, and the government duly recognized ICANN as NewCo.

ICANN's subsequent three-year history has been fraught with controversy, but only a few facets of that history need to be related: (1) The extent to which ICANN is controlled by the government, which is relevant to its status as a potential antitrust state actor; (2) the extent to which registries and registrars control ICANN and/or are controlled by it, which speaks to how they might be using ICANN to collude in anticompetitive conduct; (3) two specific instances of ICANN-imposed policies which well affect competition: the UDRP, which limits service competition among registrars, and the means by which ICANN has constrained the introduction of new TLDs, which affects competition between registries.

### 1. ICANN's Relationship with the Federal Government

ICANN is formally independent of the federal government. Its 18-person Board of Directors currently consists of four hold-over directors from the original nine self-selected incorporators, five directors elected in a somewhat controversial process[25] from each of five geographical world regions, and nine directors appointed by three different functional groups established by ICANN on corporatist lines.[26] Other than having anointed ICANN as its DNS representative, and thus approving the original incorporators, the US government has had no formal input into the selection of ICANN's directors. Like some thirty-plus other governments, the US government participates in the quarterly meetings of ICANN's government advisory committee (GAC). ICANN's rules give the GAC a right of consultation and a right to advise, but do not require ICANN to follow its instructions.[27]

---

[20] *See, e.g.*, *The Management of Internet Names and Addresses: Intellectual Property Issues--Final Report of the WIPO Internet Domain Name Process*, & 23, World Intellectual Property Organization, *at* http://wipo2.int/process1/report/finalreport.html (noting the existence of "a number of predatory and parasitical practices that have been adopted by some . . . includ[ing] the deliberate, bad faith registration as domain names of well-known and other trademarks in the hope of being able to sell the domain names to the owners of those marks"); see generally Jessica Litman, *The DNS Wars: Trademarks and the Internet Domain Name System*, 4 **J. Sm. & Emerging Bus. L.** 149 (2000).

[21] *See* Angela Proffitt, *Drop the Government, Keep the Law: New International Body for Domain Name Assignment Can Learn from United States Trademark Experience*, 19 LOY. L.A. ENT. L.J. 601, 608 (1999) (noting the concerns of the European Union, the Australian government, and others that the United States had "too much control over the DNS").

[22] Management of Internet Names and Addresses, 63 Fed. Reg. 31,741 (1998), http://www.ntia.doc.gov/ntiahome/domainname/6_5_98dns.htm [hereinafter White Paper].

[23] 31 U.S.C. § 9101-10 (1994).

[24] *See* White Paper at 31,744.

ICANN's direct relationship with the U.S. government is defined by five elements: The White Paper, three separate government contracts, and a less formal but highly significant component made up of DoC's recognition of ICANN as the entity contemplated in the White Paper plus whatever continuing supervision the Department of Commerce (DoC) may choose to exercise over its contractor.

The White Paper. Although, as a mere policy statement, the White Paper had no direct legal force, its very vagueness on key points makes it the closest thing to a consensus document produced on DNS matters in the last four years. In June 1999, ICANN stated that the White Paper "principles ... have dictated ICANN's policy decisions to date."[28] More recently, DoC reaffirmed the White Paper's centrality to DNS policy,[29] as has ICANN's Vice President and General Counsel.[30] The White Paper instructed ICANN to undertake specific tasks, including fostering competition among registrars, and attacking the cybersquatting issue.[31] From its creation as the body seeking to be anointed as NewCo to present day, ICANN has assiduously undertaken to accomplish each of the specific goals set out in the White Paper.

The White Paper opined that NewCo "should be headquartered in the United States, and incorporated in the U.S. as a not-for-profit corporation. It should, however, have a board of directors from around the world"[32] and ICANN complied. The White Paper said that NewCo should take over the existing IANA staff, and ICANN, with DoC's cooperation, later did just that. NewCo, said the White Paper, should have the authority to "[s]et policy for and direct allocation of IP number blocks to regional Internet number registries" and "[o]versee operation of the authoritative Internet root server system"[33] plus [o]versee policy for determining the circumstances under which new TLDs are added to the root system" while coordinating "the assignment of other Internet technical parameters as needed."[34] If DoC's contracts with ICANN did not necessarily give it this authority directly, they created the conditions in which ICANN could, with DoC's at least tacit blessing, exercise it for all practical purposes.

---

[25] *See* Weinberg, *supra* note 5.

[26] *Id.* There is actually a 19th director: the ICANN CEO serves *ex officio*.

[27] ICANN, Bylaws For Internet Corporation For Assigned Names And Numbers, http://www.icann.org/general/bylaws.htm. (Amended July 16, 2000).

[28] *Status Report to the Department of Commerce*, § IV, *at* http://www.icann.org/ statusreport-15june99.htm (June 15, 1999) § I.

[29] *See* Letter from John F. Sopko, Acting Assistant Secretary for Communications and Information to William F. Bode, June 25, 2001 (reiterating White Paper's "recogni[tion] that the selection of new TLDs should be conducted by the private sector through a not-for-profit organization, globally representative of the Internet stakeholder community." http://www.icannwatch.org/article.php?sid=237

[30] http://www.dnso.org/clubpublic/council/Arc05/msg00613.html

[31] It also set out four general goals for a the non-profit entity that was to manage the DNS: "stability, competition, private bottom-up coordination, and representation". ICANN wasted no time addressing the specific tasks in the White Paper, but some of these more general goals have proved more difficult to achieve, especially they conflict with each other. With "stability" listed in the White Paper as "the first priority of any DNS management system." ICANN has argued that some of the other goals, notably representation, needed to take a back seat.

The White Paper's specific policy directions included a requirement that NewCo require specified information about domain name registrants be included in all registry databases and freely available on the Internet in order to allow trademark holders to "contact a domain name registrant when a conflict arises between a trademark holder and a domain name holder."[35] Registrants should be required to pay fees at the time of registration, and required to "agree to submit infringing domain names to the authority of a court of law in the jurisdiction in which the registry, registry database, registrar, or the 'A' root servers are located."[36] NewCo should also require registrants to agree to arbitration in cases of alleged cybersquatting[37] and give special protections for famous trademarks.[38] With the exception of the special protection for famous marks, which foundered on an inability to agree on how to identify which marks were sufficiently famous, ICANN quickly implemented each of these directives.

The White Paper also prescribed a structure for NewCo's board of directors. The Board "should be balanced to equitably represent the interests of IP number registries, domain name registries, domain name registrars, the technical community, Internet service providers (ISPs), and Internet users (commercial, not-for-profit, and individuals) from around the world"[39] but government officials would be forbidden to serve on the board. The interim board would "develop policies for the addition of TLDs, and establish the qualifications for domain name registries and domain name registrars within the system."[40] ICANN faithfully followed most of these directions, although it took a long time to elect user representatives, and opinions differ on whether even today users and non-profits are equitably represented on the Board.

---

[32] *Id.* at 31,750. While the White Paper itself does not use the name "NewCo," the use of the term by DoC to describe the entity called for in the White Paper dates at least from *Cooperative Agreement No. NCR-9218742, Amendment 11*, http://www.networksolutions.com/en_US/legal/internic/cooperative-agreement/amendment11.html (Oct. 6, 1998).

[33] White Paper, *supra* note 22, at 31,749.

[34] *Id.*

[35] *Id.*

[36] *Id.*

[37] *Id.*

[38] *Id.* at 31,751.

[39] *Id.* at 31,750.

[40] *Id.*

The Three Contracts. Formally, the federal government administers its relationship with ICANN via three agreements: (1) a Memorandum of Understanding,[41] (2) a Cooperative Research and Development Agreement (CRADA);[42] and (3) an unusual no-cost, no-bid "procurement" contract for the "IANA function".

*The MoU*. The MoU was DoC's first agreement with ICANN, signed even before DoC recognized ICANN as NewCo. DoC and ICANN agreed to "jointly design, develop, and test the mechanisms, methods, and procedures that should be in place and the steps necessary to transition management responsibility for DNS functions now performed by, or on behalf of, the U.S. Government to a private-sector not-for-profit entity" in order to prepare the ground for the transition of DNS management to ICANN.[43] The "DNS management functions" included oversight of both "the operation of the authoritative root server system" and "the policy for determining the circumstances under which new top-level domains would be added to the root system," plus any other agreed activities "necessary to coordinate the specified DNS management functions."[44] Echoing the White Paper, the DoC-ICANN MoU also listed four principles by which the parties "will abide": stability of the Internet; competition; private, bottom-up coordination; and representation.[45] The MoU appears to authorize no more than a study of how the DNS would be privatized in the future. In fact, however, the DoC-ICANN MoU conveyed very significant authority, because the means by which ICANN would "study" the future privatization of the DNS was by acting as if the DNS were already privatized.[46]

A year later, DoC and ICANN amended the MoU. ICANN promised not to amend its standard

---

[41] Memorandum of Understanding, Dept. of Commerce and ICANN, at http://www.icann.org/general/icann-mou-25nov98.htm (Nov. 25, 1998) [hereinafter Memorandum of Understanding].

[42] Cooperative Research & Development Agreement, at http://www.icann.org/committees/dns-root/crada.htm [hereinafter CRADA].

[43] *Id*. § II.B.

[44] *Id*.

[45] *Id*. § II.C.

[46] As DoC later explained to a House Committee,

ICANN's responsibility under the [MoU] is to act as the not-for-profit entity contemplated in the White Paper, and to demonstrate whether such an entity can implement the goals of the White Paper. If it cannot, Government involvement in DNS management would likely need to be extended until such time as a reliable mechanism can be established to meet those goals. The Department does not oversee ICANN's daily operations. The Department's general oversight authority is broad, and, if necessary, the Department could terminate the agreement and ICANN's role in this aspect of DNS management with 120 days notice.

Letter from Andrew J. Pincus, DoC General Counsel, to Rep. Tom Bliley, Chairman, United States House Committee on Commerce (July 8, 1999), National Telecommunications and Information Administration, http://www.ntia.doc.gov/ntiahome/domainname/blileyrsp.htm .

form agreement with Registries without DoC's prior approval. ICANN also promised not to make agreements with a successor registry without DoC's approval and to follow DoC's lead if it chose to replace NSI with a new registry. And most importantly (in that it gave DoC additional leverage over ICANN), ICANN agreed that "[i]f DOC withdraws its recognition of ICANN or any successor entity by terminating this MOU, ICANN agrees that it will assign to DOC any rights that ICANN has in all existing contracts with registries and registrars."[47] Whether this "termination" language would apply if the agreement were allowed to lapse instead of being actively ended by DoC is an interesting question; the ambiguity may give ICANN leverage in any contract negotiations.

*The IANA Procurement*. DoC issued a sole source contract to ICANN for the IANA function on the grounds that ICANN was the only responsible source available.[48] DoC duly issued a purchase order to ICANN for IANA services, a purchase order that has a price of zero dollars but allows ICANN to establish and collect fees from third parties, subject to review by DoC, so long as the fees reflect the actual cost of providing the service.[49]

*The CRADA*. In the ICANN-DoC MoU, the parties had agreed that ICANN would study the privatization of the DNS by doing it. However, IANA, a separate government contractor, was already doing the job that ICANN proposed to privatize.[50] In the June 1999 Cooperative Research and Development Agreement (CRADA)[51] DoC engaged ICANN to study how to improve the IANA functions. Again, like the ICANN-DoC MoU, this new agreement appears to include having ICANN perform the

---

[47] *ICANN/DOC Memorandum of Understanding, Amendment 1*, *at* http://www.icann.org/nsi/amend1-jpamou-04nov99.htm (Nov. 4, 1999)

[48] Letter from Robert P. Murphy, General Counsel, General Accounting Office, to Sen. Judd Gregg, Chairman, United States Senate Subcommittee on Commerce, Justice, State, and the Judiciary 25 (July 7, 2000), GAO.OGC—00-33R, "Commerce and ICANN", http://www.gao.gov/new.items/ og00033.pdf, at 17.

[49] *See Contract Between ICANN and the United States Government for Performance of the IANA Function*, *at* http://www.icann.org/general/icann-contract-09feb00.htm (showing a copy of the purchase order). DoC later extended the "purchase order" for one year. See *Announcement: ICANN and U.S. Government Agree to Extend Agreements*, *at* http://www.icann.org/announcements/icann-pr04sep00.htm (Sept. 4, 2000). This extension affected both the ICANN-DoC MoU of November 25, 1998, and ICANN's Cooperative Research and Development Agreement, *see Cooperative Research & Development Agreement*, *at* http://www.icann.org/committees/dns-root/crada.htm

[50] A copy of what appears to be an agreement between that contractor, USC, and ICANN, dated January 1, 1999, appears as Appendix 21 to ICANN's application for tax-exempt status. *See Form 1023 (Appendix 21)*, *at* http://www.icann.org/financials/tax/us/appendix-21.htm (last modified Sept. 4, 2000); *see also Form 1023 (Appendix 19)*, *at* http://www.icann.org/financials/tax/us/appendix-19.htm (last modified Sept. 4, 2000) (detailing a loanout agreement for two employees)

[51] A CRADA is usually an agreement in which, as the United States Geological Survey explained:
[t]he collaborating partner agrees to provide funds, personnel, services, facilities, equipment or other resources needed to conduct a specific research or development effort while the Federal government agrees to provide similar resources *but not funds* directly to the partner.... The CRADA vehicle provides incentives that can help speed the commercialization of Federally-developed technology, making it an excellent technology transfer tool.
*What Is a CRADA?*, United States Geological Survey, *at* http://www.usgs.gov/tech-transfer/ what-crada.html (last modified May 23, 1997).

function during the study.[52]

        <u>Ongoing supervision</u>  DoC's supervision of ICANN has three visible forms.

        First, DoC's contracts with ICANN and NSI (VeriSign) require NSI to secure written instructions from DoC before making changes in the root file.[53] Thus, any of ICANN's recommendations on new TLDs require at least rubber-stamp approval from DoC; while this might be thought to provide an occasion for review it has not in fact done so. Indeed, in a recent letter denying a petition for rulemaking on the subject of new gTLDs, DoC reiterated that, following the White Paper, DoC would as a matter of policy approve ICANN's decisions without subjecting them to review.[54]

        Second, DoC's MoU with ICANN provides for oversight and cooperation, although from the outside it often is difficult to tell how much of this there is at any given moment. ICANN sends DoC an annual report about its performance under the MoU.[55] DoC promised to devote more than a quarter of a million dollars in staff time[56] and expenses to monitoring and helping ICANN. DoC's accounts of the actual intensity of this effort have varied. A DoC official testified that "[t]he Department's general oversight under the joint project is limited to ensuring that ICANN's activities are in accordance with the joint project MOU, which in turn requires ICANN to perform its MOU tasks in accordance with the White Paper."[57] But, when pressed for specifics, DoC stated that it "consults" with ICANN before its major decisions, such as ICANN's proposal to charge a fee of $1 per domain name.[58] DoC clearly supported ICANN during its first thirteen months by pressuring its other contractor, NSI, to recognize ICANN.[59] Since then, it has

---

[52] *See* CRADA, *supra* note 42; GAO Report, *supra* note 48, at 18.

[53] *See* Cooperative Agreement No. NCR-9218742, Amendment 11, http://www.networksolutions.com/legal/internic/cooperative-agreement/amendment11.html (Oct. 6, 1998).

[54] "In July 1998, the Department of Commerce made it clear that it would not participate in the selection process of new TLDs as set forth in the Statement of Policy, entitled Management of Internet Names and Addresses, 63 Fed. Reg. 31741 (1998). In the Statement of Policy, the Department recognized that the selection of new TLDs should be conducted by the private sector through a not-for-profit organization, globally representative of the Internet stakeholder community. The Department recognized ICANN as that organization in November 1998 through a Memorandum of Understanding." Letter from John F. Sopko, Acting Assistant Secretary for Communications and Information to William H. Bode (June 25, 2001), reprinted at http://www.icannwatch.org/article.php?sid=237.

[55] First status report; Second Status Report Under ICANN/US Government Memorandum of Understanding (30 June 2000), http://www.icann.org/general/statusreport-30jun00.htm (June 30, 2000); Third Annual Status Report to the US Department of Commerce, http://www.icann.org/general/statusreport-03jul01.htm

[56] DoC promised half-time dedication of four or five full-time employees. See Memorandum of Understanding, *supra* note 41, at 7.

[57] Pincus, *supra* note 43.

[58] *See id*.; *see also* GAO Report, *supra* note 48, at 23 (discussing the cooperation between ICANN and DoC regarding the above-mentioned fee).

[59] *See*, for example, DoC's statement that:
Network Solutions has indicated that it is not obligated to enter into a contract with ICANN because the Department been harder to gauge the intensity of the department's oversight, barring the occasional extraordinary intervention into ICANN's affairs. Indeed, DoC has occasionally intervened publicly in ICANN's affairs, such as its month-long review and ultimate amendment of ICANN's proposed renegotiation of its agreement with VeriSign, undertaken after some prodding by Congress.[60] Most recently, DoC Secretary Donald Evans wrote to ICANN urging it to approve more top-level domain names soon,[61] although it is not clear if this letter had, or even was intended to have, any direct effect.

Perhaps most important, but least visible, is the sword of Damocles that DoC holds over ICANN's head. ICANN's powers stem from its contracts, and its recognition by DoC as the "'NewCo" specified in the White Paper. The original MoU allowed DoC to terminate the agreement on September 30, 2000, or after 120 days' notice, and subsequent extensions have had similar provisions.[62] As all the DoC-ICANN contracts run require annual or semi-annual renewal, they provide a means for DoC to pressure ICANN were it to choose to do so. In theory, DoC could transfer its imprimatur from ICANN to another body, although in practice this seems a remote possibility.

### 2. Registries' and Registrars' Structural Relationships With ICANN

Adhering, more or less, to the White Paper's directions regarding the internal organization of NewCo,[63] ICANN created three subsidiary councils charged with developing policy and making recommendations to the Board. Each of these three groups also elected three of the ICANN Board's 18 directors. One of these three bodies, the Domain Name Supporting Organization (DNSO) is charged with concentrating on domain name related issues. The DNSO gives registrars (and, now that there are more than one, gTLD registries) a place to meet, to lobby ICANN, and to exert some influence on the selection of its directors.

The DNSO is sub-divided into seven 'stakeholder' constituencies selected by the nine initial self-selected ICANN directors: registrars,[64] gTLD registries, ccTLDs, ISPs, trademark holders, businesses,

---

of Commerce has not "recognized" ICANN by transferring authority over the authoritative root system to it. We find no merit in this argument. The Department of Commerce entered into a Memorandum of Understanding with ICANN on November 25, 1998. That MOU constitutes the Government's "recognition" of ICANN. We reiterated this point in a letter to Network Solutions on February 26, 1999.

Pincus, *supra* note 43.

[60] http://www.icannwatch.org/article.php?sid=159; Commerce Ensures Competitiveness and Stability are Protected in New ICANN-VeriSign Agreement, http://osecnt13.osec.doc.gov/public.nsf/docs/icann-verisign-0518; Statement by Department of Commerce General Counsel Ted Kassinger Regarding the Proposed Verisign-Icann Agreement, http://osecnt13.osec.doc.gov/public.nsf/docs/Kassinger-on-ICANN-Verisign

[61] Letter from Donald L. Evans, U.S. Secretary of Commerce, to Vint Cerf (25 May 2001), http://www.icann.org/correspondence/doc-to-icann-25may01.htm

[62] *See* Memorandum of Understanding, *supra* note *, ' VII.

[63] *See* White Paper, *supra* note 22.

[64] Oddly, future registrars were allowed to enter the registrars' constituency and vote before they were accredited by ICANN, but future registries were not.

and non-commercial domain name holders.[65] Each of the seven constituencies elects three[66] representatives to the DNSO's governing body, the Names Council, which in turn elects three representatives to the ICANN Board. The registrars' strength may be greater than it seems, however, since registrars could join more than one constituency simultaneously. Indeed, a single firm could simultaneously be a member of four: commercial and business entities; ISP and connectivity providers; registrars; and trademark, other intellectual property, and anti-counterfeiting interests.[67] Thus, in theory, registrars could influence a majority of the votes on the Names Council, which in turn might allow them to choose three of ICANN's directors.

ICANN has substantial power over registrars, as its non-negotiable standard form Registrar Agreement requires them to pledge to observe ICANN's policy decisions,[68] and also gives ICANN the power to disqualify a registrar.[69] gTLD Registries must make a similar pledge to follow ICANN's consensus policies, but as their contracts are each unique, ICANN also is able to impose additional requirements on the new gTLDs before allowing them to join the legacy root. ICANN's control does not (yet) extend to the ccTLD registries, although ICANN is currently negotiating agreements with them as well.

ICANN's control over the registrars stems in part from its agreements with the registries. In particular, ICANN's agreement with NSI, the monopoly gTLD registry until the forthcoming introduction of .biz, .info and other new gTLDs, requires NSI to ensure that registrars accept ICANN's standard form Registrar Agreement,[70] before allowing them to register any names. And the chief consensus policy grandfathered into the NSI Registry Agreement is ICANN's mandatory arbitration clause for domain name disputes, the Uniform Dispute Resolution Policy (UDRP).[71]

---

[65] *See* The seven constituencies official names are: ccTLD registries; commercial and business entities; gTLD registries; ISP and connectivity providers; noncommercial domain name holders; registrars; and trademark, other intellectual property, and anti-counterfeiting interests. *See About DNSO*, Domain Name Supporting Organization, *at* http://www.dnso.org/dnso/aboutdnso.html; cf. Jonathan Weinberg, *ICANN and the Problem of Legitimacy*, 50 DUKE L.J. 187, 238 & n.261 (2000).

[66] ICANN stripped one of the seven DNSO constituencies,, the gTLD constituency, of two of its three Names Council Representatives because there was only one firm, NSI/VeriSign, represented in the constituency. The full three-member representation is due to be restored when new gTLD registries join the constituency.

[67] *See By-Laws of the Intellectual Property Constituency*, § III, Intellectual Property Constituency, *at* http://ipc.songbird.com/IPC_Bylaws_dec_15_correct.htm (Nov. 30, 1999); *Commercial and Business Entities Constituency Charter*, § II.A, Business Constituency Domain Name Supporting Organization, *at* http://www.bcdnso.org/Charter.htm (Oct. 25, 1999); *The DNSO Registrar Constituency*, § II, Domain Name Supporting Organization, *at* http://www.dnso.org/constituency/registrars/Registrars.Articles.html; *ISPs and Connectivity Providers, How to Become a Member*, Domain Name Supporting Organization, *at* http://www.dnso.org/constituency/ispcp/membership.html.

[68] *Registrar Accreditation Agreement*, http://www.icann.org/nsi/icann-raa-04nov99.htm (Nov. 4, 1999)

[69]' ICANN, Statement Of Registrar Accreditation Policy, § II.C. http://www.icann.org/policy_statement.html#IIC

[70] *See* ICANN-NSI Registry Agreement, http://www.icann.org/nsi/nsi-registry-agreement-04nov99.htm (Nov. 4, 1999); *Registrar Accreditation Agreement*, *at* http://www.icann.org/nsi/icann-raa-04nov99.htm

[71] ICANN, Registrar Accreditation Agreement, http://www.icann.org/nsi/icann-raa-04nov99.htm § II.K (Nov. 4, 1999).

### 3. ICANN Policies With Competitive Implications

(a) UDRP

The White Paper recommended that NewCo require registrants to agree "that in cases involving cyberpiracy or cybersquatting (as opposed to conflicts between legitimate competing rights holders), they would submit to and be bound by alternative dispute resolution systems identified by the new corporation for the purpose of resolving those conflicts."[72] The White Paper itself said little about what this dispute resolution policy should look like, choosing instead to ask the World Intellectual Property Organization (WIPO), to advise NewCo on a plan. WIPO duly did just that,[73] and once it had WIPO's advisory report in hand ICANN wasted no time in starting its cumbersome policy-making machinery to address this issue. A "working group" was formed to consider the issues, and eventually recommended that some sort of WIPO-like policy was appropriate.[74]

Meanwhile, however, the registrars became impatient with the slow rate of progress of the working group process. In 1999, ICANN had begun accrediting new registrars who wished to compete with NSI. Although ICANN initially proposed a fairly detailed intellectual property protection regime,[75] in March, 1999 the Board adopted a policy that mostly put the question off until it decided what it should do with WIPO's recommendations.[76]

As they prepared to go live in the so-called 'testbed' phase in mid-1999, the newly accredited registrars found themselves in a delicate position. On the one hand, the trademark interests were telling them that they faced exposure to liability if they didn't institute some method of protecting trademark holders against cybersquatters. On the other hand, the policy then used by NSI was obviously draconian and unfair.[77] The registrars, perhaps with ICANN's encouragement, decided to draft their own dispute policy. By May, 1999 the ICANN staff reported to the Board that "the [Registrar Accreditation] Agreement calls for registrars to adopt dispute resolution policies, and that accredited registrars are already working together to do so. Counsel noted that Network Solutions' registry-registrar contract also calls for registrars to have dispute resolution policies in place, and that registrars are anxious to have guidance on a uniform policy."[78]

---

[72] White Paper, *supra* note 22, at 31,750.

[73] WIPO, *The Management of Internet Names and Addresses: Intellectual Property Issues: Final Report of the WIPO Internet Domain Name Process*.

[74] Working groups are supposed to be open to all comers; in fact, however, at least one of these groups was manipulated to exclude opponents of mandatory arbitration. See A. Michael Froomkin, *Comments on ICANN Uniform Dispute Policy: A Catalog of Critical Process Failures; Progress on Substance; More Work Needed*, at http://www.law.miami.edu/~amf/icannudp.htm (Oct. 13, 1999).

[75] In the Proposed Accreditation Guidelines for Registrars, ICANN stated that the final document "should protect legal rights (including intellectual property rights) of the parties, and of third parties where applicable. It should contain provisions that minimize disputes over rights to use of particular domain names, and in the event of dispute, it should contain provisions that enhance the orderly and timely resolution of disputes." http://www.icann.org/singapore/draftguidelines.htm at § I.D.3 (February 8, 1999). Section K of this document also contained a list of WIPO recommendations that ICANN thought should be incorporated into the Registries' practices.

[76] "During the term of the accreditation agreement, the registrar will have in place a policy and procedure for resolution of disputes concerning SLD names. In the event that ICANN establishes a policy or procedure for resolution of disputes concerning SLD names that by its terms applies to the registrar, the registrar will adhere to the policy or procedure." ICANN, STATEMENT OF REGISTRAR ACCREDITATION POLICY (.com, .net, and .org top-level domains) (Adopted March 4, 1999) http://www.icann.org/policy_statement.html#IIIK.
This clause became § III.J. of the Registrar Accreditation Agreement, http://www.icann.org/registrars/ra-agreement-12may99.htm.

The Registrars' desire to have a tough and uniform policy was exacerbated by political and competitive factors. By the time the new registrars entered the scene, the conventional wisdom increasingly was that 'the best names are taken'. The registrars as a group were therefore desperately anxious to have new product to sell - registrations in new gTLDs in which they would be competing more evenly with NSI. They also believed, not without reason, that trademark owners had a virtual veto over the creation of new gTLDs, and that they would exercise it until they were satisfied by the intellectual property protections instituted by the registrars. This created a powerful incentive to draft a tough policy on cybersquatting in order to placate the trademark owners.

Tough wasn't enough. In order to satisfy the trademark interests the policy also had to be uniform - to apply to all accredited registrars, and on all registrants. Indeed, the registrars themselves had a vested interest in ensuring that the policy applied to all their competitors, and especially their competitors' customers, else they feared some registrar might compete on service terms and attract disproportionate business by being 'registrant friendly' -- or worse, 'cybersquatter friendly'. This fear was far from academic, as prior to ICANN's UDRP, registrars had a variety of policies in place.[79]

Even before the first ICANN working group on domain name arbitration reported in late July 1999,[80] a group of registrars[81] quietly employed a Skadden Arps lawyer, Rita Rodin, to craft a dispute policy for them. On August 20, 1999, the registrars unveiled their proposed policy document.[82] Four days later, the ICANN staff issued a report with its own detailed suggestions about what the dispute policy should look like, many of which followed the registrars' lead.[83] Two days after that, amidst much controversy, the ICANN Board was resolved to use the registrars' draft, rather than anything drafted through the ICANN consensus policymaking procedure "as a starting point" for the drafting of ICANN's own policy.[84] In practice this meant that the Registrars' draft was accepted almost in toto, save that a few of the most controversial issues were referred to a 'small drafting committee' made up of representatives of the warring factions.[85] This committee was only advisory, and the staff did not accept all of its suggestion even when it was able to reach consensus.[86] Ultimately, ICANN staff prepared the final draft of the UDRP, a text that owed a great deal to the registrars' draft, which in turn relied on some of WIPO's suggestions.[87]

---

[77] See Carl Oppedahl, *How Is a Domain Name Like a Cow?*, 15 **J. Marshall J. Computer & Info. L.** 437 (1997).

[78] ICANN, Minutes, Meeting of the Initial Board, May 27, 1999, http://www.icann.org/minutes/berlinminutes.html

[79] *See* http://www.domainhandbook.com/dompol.html. As NSI was the registry, they were all de facto subject to its policies, although there was a legal question as to liability for NSI's actions in the absence of privity with the customer.

[80] *See* WG-A Final Report to the Names Council - July 29, 1999 - REVISED DRAFT, http://www.dnso.org/dnso/notes/19990729.WGA-report.html.

[81] By August, 1999 this group of 20 or so registrars apparently included Alibanza, AOL, AT&T, AITcom, CORE, Domainbank, FICPI, Infonetworks, Interq-Japan, Netnames, NSI, PSI Japan, Register.com plus the ICC and INTA. *See* POLICY STATEMENT REGARDING THE MODEL DOMAIN NAME DISPUTE POLICY, http://www.dnso.org/constituency/registrars/Website/udrp-draft-19990909.html. The official Registrars' Statement offered a slightly different list of Participating Registrars, Alabanza, Inc., America Online Incorporated, Animus Communications, Inc., Domain Bank, Inc., EnetRegistry.com Corporation, eNOM, Inc., InfoNetworks, Inc., Melbourne IT, Network Solutions, Inc., Nominalia Internet S.L., register.com, Tech Dogs, Inc., TUCOWS.com, Inc., and WebTrends Corporation. ICANN, Registrars' Statement Regarding Their Model Uniform Dispute Policy (Posted August 20, 1999), http://www.icann.org/santiago/registrar-policy-statement.htm.

(b) Constrained Roll-out of New TLDs

---

[82] The Registrars' policy was unveiled on Aug 20, and voted on by the ICANN Board on five days later. *See* ICANN Staff Report: Uniform Dispute Resolution Policy for gTLD Registrars, http://www.icann.org/santiago/udrp-staff-report.htm.

[83] ICANN Staff Report: Uniform Dispute Resolution Policy for gTLD Registrars, http://www.icann.org/santiago/udrp-staff-report.htm (Aug. 24, 1999).

[84] See Resolution 99.83(1), http://www.icann.org/santiago/santiago-resolutions.htm.

[85] The Board resolution, id., stated:
  FURTHER RESOLVED [99.82] that the President is directed, with the assistance of ICANN staff and counsel, to prepare implementation documents for approval by the Board after public notice and comment, on a schedule that allows the policy to be put into place within 45 days.
  FURTHER RESOLVED [99.83] that the Board gives the following guidance as to the preparation of the implementation documents:
  1. The registrars' Model Dispute Resolution Policy should be used as a starting point;
  2. The President or his delegate should convene a small drafting committee including persons selected by him to express views and consider the interests of the registrar, non-commercial, individual, intellectual property, and business interests;
  3. In addition to the factors mentioned in paragraph 171(2) of the WIPO report, the following should be considered in determining whether a domain name was registered in bad faith:
     (a) Whether the domain name holder is making a legitimate noncommercial or fair use of the mark, without intent to misleadingly divert consumers for commercial gain or to tarnish the mark
     (b) Whether the domain name holder (including individuals, businesses, and other organizations) is commonly known by the domain name, even if the holder has acquired no trademark or service mark rights; and
     (c) Whether, in seeking payment for transfer of the domain name, the domain name holder has limited its request for payment to its out-of-pocket costs.
  4. There should be a general parity between the appeal rights of complainants and domain name holders.
  5. The dispute policy should seek to define and minimize reverse domain name hijacking.

[86] *See Second Staff Report on Implementation Documents for the Uniform Dispute Resolution Policy*, at http://www.icann.org/udrp/udrp-second-staff-report-24oct99.htm (Oct. 24, 1999).

[87] *Id.* For a detailed discussion of the UDRP's genesis and content see A. Michael Froomkin, *ICANN's "Uniform Dispute Resolution Policy"--Causes and Partial Cures* (forthcoming), http://www.law.miami.edu/~froomkin/articles/udrp.pdf.

ICANN selected the first seven new gTLDs for inclusion in the legacy root at its second Annual Meeting in Los Angeles from a crowded and highly contentious field of 47 applicants, each of whom had paid a non-refundable $50,000 fee to have their applications considered.[88] In one sense, the November, 2000 decision was the culmination of almost two years of effort; in another, it was only the start of another nine or more month of tough bargaining over the contract terms that would bind each registry to ICANN, some of which continues currently. After each contract is painstakingly negotiated, ICANN submits the gTLD to DoC; the first approvals happened within a few hours of the submission, suggesting that DoC's review was somewhat cursory.[89]

Breaking the logjam that had prevented any new gTLDs from joining the root[90] was of course of the main reasons why DoC wanted ICANN to exist, and why it contracted with it. ICANN's internal processes leading up to the selection of new gTLDs reflected the divisions in the various affected communities, the details of which need not concern us here. At no time prior to its decision to approve only a limited number of new TLDs did ICANN issue an opinion explaining the technical justification for this (or any other) limit. Nor did ICANN refer to such a report by anyone else. In fact, so far was we can discern, no such study, report or analysis existed then, or exists now. ICANN's decision was fundamentally political: an ICANN working group brokered a deal between the faction that wanted a very large number of new TLDs, and those who wanted none. In April 1999, the DNSO Names Council voted to "recommend to the Board that a limited number of new top-level domains be introduced initially and that the future introduction of additional top-level domains be done only after careful evaluation of the initial introduction."[91] In so doing, it endorsed the recommendation of that Working Group, which had compromised on "six to ten, followed by an evaluation period."[92]

ICANN's decision to limit the number of new gTLDs to well below the lowest estimates of what the DNS could handle reduced the possible competition between registries. ICANN justified it on the grounds of compromise, but also on the grounds that it had been a long time since a new gTLD had been introduced, and there might therefore be Internet 'stability' issues to consider that required a 'test' or 'proof

---

[88] http://www.icann.org/minutes/prelim-report-16nov00.htm#SecondAnnualMeeting.

"The selected TLD proposals are of two types. Four proposals (.biz, .info, .name, and .pro) are for relatively large, unsponsored TLDs. The other three proposals (.aero, .coop, and .museum) are for smaller "sponsored" TLDs. (Generally speaking, an "unsponsored" TLD operates under policies established by the global Internet community directly through the ICANN process, while a "sponsored" TLD is a specialized TLD that has a sponsoring organization representing the narrower community that is most affected by the TLD". http://www.icann.org/tlds/ .

[89] *See* http://www.icannwatch.org/article.php?sid=222; see also supra note (discussing DoC letter explaining that it follows White Paper in leaving gTLD decisions to ICANN).

[90] In contrast to gTLDs, adding ccTLDs appeared to be uncontroversial. For example, the US Dept. of Commerce authorized the creation of the .ps. ccTLD after receiving ICANN's recommendation that it do so, delivered in ICANN, IANA Report on Request for Delegation of the ps Top-Level Domain, http://www.icann.org/general/ps-report-22mar00.htm (Mar. 22, 2000).

[91] http://www.icann.org/yokohama/new-tld-topic.htm#IC .

[92] Report (Part One) of Working Group C (21 March 2000), http://www.icann.org/dnso/wgc-report-21mar00.htm.

of concept' period. Although one might have doubts about the validity of this technical rationale, since new ccTLDs had been introduced without any noticeable effect on anyone, there can be no doubt that the method ICANN chose to select the TLDs substantially reduced competition in other ways that had no technical justification.[93] There were, however, two ways in which the introduction of new gTLDs genuinely would be new. First, there was a huge pent-up demand for 'good' domain names, leading to fears of a chaotic 'landrush' period in the early moments of any new registry. Second, there was an exquisitely heightened sensitivity to the concerns of trademark holders who believed not only that that they should be protected from a fresh round of cybersquatting, but that trademark owners ought to be first in the queue for new names. The argument that the decision to introduce only a small number of new gTLDs and to make no promises about when if ever there would be more actually increased the likelihood of cybersquatting, since it failed to increase the supply of new names to the point where there was no likelihood of profit from hoarding failed to persuade either the markholders' representatives or ICANN. Instead, ICANN justified the small number of gTLDs as a cautious reaction to uncertainty in light of the Internet's vastly increased size and commercial importance,[94] a view that echoed the policy direction in the White Paper.[95]

---

[93] Among these were ICANN's decision to require a non-refundable $50,000 "application fee," ICANN's requirement that successful applicants demonstrate huge financial reserves; ICANN's decision to have most new TLDs limited by restrictive charters rather than being able to sell domains to all comers; and ICANN's decision to select the names of the new gTLDs itself rather than letting the winners do it on the basis of their market research.

[94] Probably the clearest, and yet very carefully nuanced, statement of this view came after the fact from Vinton Cerf:
> Of course, it cannot be stressed enough that no one knows for sure what the effects of this experiment will be. Since there have been no new global TLDs introduced for more than a decade, the Internet is a vastly different space than it was the last time this happened. Of course, there have been a number of country code TLDs introduced over that period, and since some of those have recently begun to function in a way quite analogous to a global TLD, it may be that we will be able to conclude that the DNS can readily absorb more new global TLDs. But there has never been an introduction of as many as seven new global TLDs simultaneously, with the possibility of a land rush that is inherent in that fact. There has never been a highly visible introduction of multiple new TLDs in the context of an Internet that has become a principal global medium for commerce and communication. We do not know whether the introduction of a number of new TLDs -- especially combined with the relatively new phenomenon of the use of ccTLDs in a fashion never intended (after all, .tv stands for Tuvalu, not television, no matter what its marketers say) -- will create consumer confusion, or will impair the functioning of various kinds of software that has been written to assume that .com is the most likely domain for any address.
>
> In short, it is not absolutely clear what effects these introductions will have on the stability of the DNS or how to introduce new TLDs in a way that minimizes harmful side-effects, and that is precisely why we are conducting this experiment. The results will guide our future actions.

Testimony of Vint Cerf Before U.S. House Committee on Energy and Commerce, Subcommittee on Telecommunications and the Internet (Feb. 8 2001).

[95] "At least in the short run, a prudent concern for the stability of the system suggests that expansion of gTLDs proceed at a deliberate and controlled pace to allow for evaluation of the impact of the new gTLDs and well-reasoned evolution of the domain space. New top level domains could be created to enhance competition and to enable the new corporation to evaluate the functioning, in the new environment, of the root server system and the software systems that enable shared registration." White Paper, *supra* note 22.

## II. Antitrust Immunity for State Action

### A. The State Action Doctrine

Government agencies and those they authorize to act are immune from antitrust scrutiny. The principal source of such immunity is the "state action doctrine," which strictly speaking protects only actions by states and those they deputize, perhaps including municipal and county governments.[96] However, there are parallel immunity doctrines protecting both the United States government and foreign sovereigns.[97]

The purpose behind the state action doctrine is the subject of some dispute.[98] It is clear that the Court views government action as different in some fundamental respect from private action, though whether the source of that difference lies in the constitutional allocation of responsibilities to the states or in the antitrust laws themselves is unsettled.[99] Regardless, the Court made it clear in *Parker v. Brown* that a state could immunize even naked private cartels from antitrust scrutiny if it were to require such anticompetitive conduct as a matter of state policy.[100] Similarly, Congress is free to exempt particular industries or kinds of conduct from the antitrust laws, so long as it does so expressly.[101] Indeed, because Congress is not subject to the dictates of the Supremacy Clause, it is free to repeal the antitrust laws themselves in whole or in part,[102] though the Court has proved reluctant to infer such a repeal in the absence of clear evidence.[103] Importantly, though, only Congress – and not federal agencies – is entitled to waive or repeal the antitrust laws.

---

[96] *See* Parker v. Brown, 317 U.S. 341 (1943); I **Areeda & Hovenkamp, Antitrust Law** ¶ 221b. The rules under which local governments are exempt under the state action doctrine are complex and not relevant here. For a discussion, see City of Lafayette v. Louisiana Power & Light, 435 U.S. 389, 412-13 (1978); Community Communications Co. v. City of Boulder, 455 U.S. 40, 50-51 (1982); I **Areeda & Hovenkamp** ¶ 223.

[97] *See* IA **Areeda & Hovenkamp** ¶ 252 (federal sovereign immunity); ¶ 274 (foreign sovereign immunity); 28 U.S.C §§ 1330, 1602-1611 (Foreign Sovereign Immunities Act).

[98] There has been a great deal of academic literature devoted to this topic. While there are many different theories, the literature might reasonably be divided into those who believe that government officials can be expected to act altruistically, and therefore don't need or deserve antitrust oversight, and those who believe government officials are subject to capture or to the dictates of public choice theory, and therefore might be expected to act anticompetitively. In the former camp, see Einer Elhauge, *Making Sense of Antitrust Petitioning Immunity*, 80 **Calif. L. Rev.** 1177 (1992); Einer Richard Elhauge, *The Scope of Antitrust Process*, 104 **Harv. L. Rev.** 667 (1991); Steven Semeraro, *Demystifying Antitrust State Action Doctrine*, 24 **Harv. J. L. & Pub. Pol'y** 203 (2000). In the latter, see Lemley & McGowan, *supra* note __; Robert P. Inman & Daniel L. Rubinfeld, *Making Sense of the Antitrust State Action Doctrine: Balancing Political Participation and Economic Efficiency in Regulatory Federalism*, 75 **Tex. L. Rev.** 1203 (1997); John S. Wiley, Jr., *A Capture Theory of Antitrust Federalism*, 99 **Harv. L. Rev.** 713 (1986); William H. Page, *Interest Groups, Antitrust, and State Regulation: Parker v. Brown in the Economic Theory of Legislation*, 1987 **Duke L.J.** 618.

[99] *See* Lemley & McGowan, *supra* note __, at __.

[100] *See Parker*, 317 U.S. at 341 (California raisin growers' cartel that destroyed 70% of its crop every year in order to "stabilize" prices was immune from antitrust scrutiny because California law authorized the cartel).

[101] *See, e.g.,* I **Areeda & Hovenkamp** ¶ 219 (exemption for insurance); ¶¶ 255-257 (labor exemption); ¶¶ 249-251 (miscellaneous other exemptions).

The paradigm case of antitrust immunity is where the government itself acts directly to restrain competition, for example by passing a law setting minimum prices or forbidding new entry into a market. Governments are themselves immune from antitrust liability in such cases, even though the restraint on competition may be quite egregious. In *City of Columbia v. Omni Outdoor Advertising*,[104] for example, the Court immunized a city council from antitrust liability for banning new billboards, even though there was good evidence in the case that the mayor and other members of the council were good friends of the existing local billboard monopolist and acted at its behest.[105]

A closer question is presented when the defendant is a private actor who claims to be acting in accordance with state policy. In both federal and state immunity cases, the question of whether a private party shares a state's immunity depends on two facts: whether the government has *clearly articulated* its intent that the private party act anticompetitively (or at least without antitrust constraint), and whether the state has *actively supervised* the subsequent conduct of the private party.[106] Only if private action is both subject to a clearly articulated government policy and actively supervised by the government will it be entitled to antitrust immunity.

Some examples may help illumine the scope of antitrust immunity for private actors. For example, in *California Retail*, the state of California had enacted a statute that protected wine dealers by authorizing them to engage in "resale price maintenance" – the practice of preventing discounting by requiring that retailers sell at no less than a certain price. Resale price maintenance is illegal per se under the federal antitrust laws,[107] and the question is whether wine dealers who engaged in such a scheme were immunized by the state statute from federal antitrust liability. In this case, the legislature was quite clear in articulating its policy.[108] Nonetheless, the Court rejected antitrust immunity because it found that the state had not actively supervised the wine dealers, but had merely delegated authority over price to them:

---

[102] *See* IA **Areeda & Hovenkamp,** *supra* note __, at ¶242d (federal immunity "quite similar" to state immunity, but federal immunity is always subject to the will of Congress, which can write immunity as broadly as it wishes).

[103] *See, e.g.,* National Gerimedical Hosp. & Gerontology Ctr. v. Blue Cross, 452 U.S. 378 (1981).

[104] 499 U.S. 365 (1991).

[105] For a discussion of the facts of the case, *see* Lemley & McGowan, *supra* note __, at 312-14.

[106] *See, e.g.,* FTC v. Ticor Title Ins. Co., cite; California Retail Liquor Ass'n v. Midcal Aluminum, 445 U.S. 97 (1980).

[107] *See*, e.g., 324 Liquor Corp. v. Duffy, 479 U.S. 335, 341-43 (1987); California Retail Liquor Dealers v. Midcal Aluminum Inc., 445 U.S. 97, 102-03 (1980); United States v. Parke, Davis & Co., 362 U.S. 29 (1960); Dr. Miles Medical Co. v. John D. Park & Sons Co., 220 U.S. 373 (1911).

[108] Indeed, the Court noted that "the legislative policy is forthrightly stated and clear in its purpose to permit resale price maintenance." *California Retail*, 445 U.S. at 105.

> The state simply authorizes price setting and enforces the prices established by private parties. The State neither establishes prices nor reviews the reasonableness of the price schedules . . . The national policy in favor of competition cannot be thwarted by casting such a gauzy cloak of state involvement over what is essentially a private price-fixing arrangement.[109]

Thus, the Court made it clear that government cannot simply abdicate its role to set and enforce policy to a private actor. To similar effect is *Ticor Title*.[110] In that case, the Federal Trade Commission alleged that six title insurance companies had conspired to fix prices. The companies defended on the grounds that they belonged to "rating bureaus" – private entities organized by the companies themselves to set uniform rates for their members – that were themselves licensed by the states and authorized to set rates, subject only to a veto by the state regulators. If the state did not object to the rate within thirty days, it took effect.[111] The Court held this scheme illegal as well. It asked whether "the State has played a substantial role in determining the specifics of the economic policy."[112] Only where "the details of the rates or prices have been established as a product of deliberate state intervention, not simply an agreement among private parties," will the Court consider the state to have actively supervised the private restraint.[113]

The rationale of these cases is clear: "Absent such a program of supervision, there is no realistic assurance that a private party's anticompetitive conduct promotes state policy, rather than merely the private party's individual interests."[114] To justify antitrust immunity, the government must have not only the right and ability to overrule private decisions, but must actually exercise its power to review those decisions.[115] And despite the Court's early deference to a price-fixing scheme in *Parker v. Brown*, the clear articulation and active supervision requirements of late have proven difficult hurdles to clear.

### B. Antitrust Immunity for Network Solutions

---

[109] *Id.* at 105-06.

[110] FTC v. Ticor Title Ins. Co., 504 U.S. 621, (1992).

[111] *Id.* at __.

[112] *Id.* at __.

[113] *Id*. The same result obtains even in the absence of pricing decisions, where the state has delegated authority over a marketplace to a private actor. *See* Patrick v. Burget, 486 U.S. 94 (1988) (peer review by Board of Medical Examiners not immune as state action, despite state authorization, because the state did not actively supervise the Board).

[114] *Patrick*, 486 U.S. at 100-01.

[115] *Id.*; *see also* FTC v. Ticor Title Ins. Co., 504 U.S. at 638; Shahawy v. Harrison, 875 F.2d 1529 (11th Cir. 1989); Pinhas v. Summit Health, 894 F.2d 1024, 1030 (9th Cir. 1989), *aff'd on other grounds* 500 U.S. 322 (1991) (both finding that deferential state review of a medical board for procedural error or arbitrary or capricious action was inadequate to clothe the board with antitrust immunity). Federal Law has since modified this rule. *See* Health Care Quality Improvement Act, 42 U.S.C. §§101-111.

Despite the rather strict set of requirements for antitrust immunity articulated in the previous section, Network Solutions, Inc. fared extremely well in antitrust litigation based on its conduct during the time before ICANN was formed. Several reported cases have considered antitrust claims against NSI; none have found liability. District courts generally rejected antitrust liability on the grounds that NSI was acting under the authority of the federal government, and so was immune from suit. The appellate courts were more cautious in granting such immunity, however. In this section, we discuss those cases, as well as two sets of related decisions, before turning in the next section to consider their implications for ICANN.

Four district courts have considered whether NSI is immune from suit because it acted at the behest of the government in setting domain name policy. All four courts concluded that NSI was immune from antitrust scrutiny, in each case applying what the courts referred to as "federal instrumentality" immunity.[116] The immunity described by these cases sweeps much more broadly than the state action antitrust immunity described in the previous section. Indeed, in *PG Media* and *Thomas*, the district courts expressly distinguished the state action cases, holding that federal immunity was broader and did not require proof of anything other than authorization pursuant to a government cooperative agreement.[117]

Courts of appeals have been much more restrictive in their reading of NSI's immunity, however. In both *Thomas* and *Watts*, the circuit courts refused to rely on immunity principles at all, instead affirming the district court decision because of another defect in the plaintiffs' antitrust cases. In *Thomas*, the court found the question of NSI's immunity "not clearly settled."[118] The court held that the United States government was clearly immune from suit under the antitrust laws.[119] By contrast, the court noted that

> It is not obvious to us . . . that a private contractor automatically shares the federal agency's immunity simply because the contractor's allegedly anti-competitive conduct occurred . . . pursuant to a government contract. A contractor might be free to perform the contract in any number of ways, only one of which is anticompetitive.[120]

---

[116] *See, e.g.,* Beverly v. Network Solutions, Inc., 1998 WL 320829, at *4 (N.D. Cal. June 12, 1998) (NSI immune because it is a private party "acting in compliance with a clearly articulated government program"; no mention of active supervision requirement); Watts v. Network Solutions, Inc., 1999 WL 778589, at *3 (S.D. Ind. May 7, 1999), *aff'd on other grounds* 202 F.3d 276 (7th Cir. 1999) (unpublished); Thomas v. Network Solutions, 2 F. Supp. 2d 22 (D.D.C. 1998), *aff'd on other grounds*, 176 F.3d 500 (D.C. Cir. 1999); PGMedia v. Network Solutions, Inc., 51 F. Supp. 2d 389 (S.D.N.Y. 1999), *aff'd sub nom. Name.space, Inc., v. Network Solutions, Inc.,* 202 F.3d 573 (2d Cir. 2000). Other cases are pending. *See, e.g.,* Chrysalis Vineyards v. U.S. Department of Commerce, No. 00-1330-A (E.D. Va. filed Sept. 7, 2000).

[117] *See Thomas*, 2 F. Supp. 2d at 38.

[118] 176 F.3d at 508.

[119] *Id*. (citing United States v. Cooper Corp., 312 U.S. 600 (1941) (U.S. government not a "person" who can be sued under the Sherman Act); Sea-Land Service v. Alaska Railroad, 659 F.2d 243 (D.C. Cir. 1981) (wholly owned and operated government corporation was immune from suit).

[120] *Id*. at 508-09 (citing Otter Tail Power Co. v. United States, 410 U.S. 366 (1973)). In *Otter Tail*, the Court seemed to reject the idea that government contracts could confer antitrust immunity, noting that "government contracting officers do not have the power to grant immunity from the Sherman Act." *Id*. at 378-79. But the court went on to suggest that

The court did not decide the issue, choosing instead to address the deficiencies it perceived in the merits of the plaintiff's antitrust claim.[121] The Seventh Circuit did the same thing in the *Watts* case. It noted that NSI's immunity was not automatic, as the district court had held, citing *Thomas*.[122] Like *Thomas*, the court chose to affirm on another antitrust ground (in that case standing) "rather than decide the complex issue of whether NSI enjoys antitrust immunity."[123] Because the *Watts* decision is unpublished, however, its endorsement of the D.C. Circuit approach is of no precedential value.

The most detailed treatment of the issue is the Second Circuit's opinion in *Name.space*.[124] In that case, the plaintiff had challenged NSI's refusal to issue new gTLDs. NSI had initially decided to issue new TLDs, but after consulting with the National Science Foundation, the NSF directed it not to do so. NSI followed this directive and refused to issue new TLDs, whereupon the plaintiff sued it for violating the antitrust laws. The district court held that NSI was entirely immune from antitrust scrutiny under the federal instrumentality doctrine.[125]

On appeal, the Second Circuit refused to apply the federal instrumentality doctrine, reasoning that "reliance on such a broad rule of immunity might improperly insulate NSI and other private entities that are or will be involved in administering the DNS from liability for future anticompetitive conduct."[126] Rather, the court applied an immunity doctrine based largely on the state action doctrine. It had little trouble finding immunity in the case before it, however, because "the conduct being challenged by Name.Space in this appeal was compelled by the explicit terms of NSI's agreement with a government agency and by the government policies."[127] To require NSI to act, yet permit it to be sued for doing what the government required it to do, would be unfair.[128]

These cases suggest that NSI's antitrust immunity for conduct related to the DNS is far from clearly settled. In fact, it is unlikely that NSI will receive absolute immunity. Rather, immunity will be determined on the basis of traditional principles drawn from the state action cases: did the government clearly articulate

---

some contracting parties may in fact be immune by virtue of their relationship to the government, leaving the Court's holding on this point less than clear.

[121] 176 F.3d at 509 (holding that Thomas had not made out the elements of an essential facilities claim).

[122] 202 F.3d 276 (7th Cir. 1999).

[123] *Id.*

[124] 202 F.3d 573 (2d Cir. 2000).

[125] PGMedia v. Network Solutions, Inc., 51 F. Supp. 2d at 407.

[126] 202 F.3d at 581.

[127] *Id.* at 582.

[128] *Id.* at 583 (quoting Alpha Lyracom Space Comm. v. Communications Satellite Corp., 946 F.2d 168, 174 (2d Cir. 1991) "Congress could not have intended to require a private entity to act subject to federal governmental directives and, at the same time, have intended that it proceed at its own antitrust peril in carrying out that official role.").

a policy that requires interference with competition, and did the government actively supervise private decision-making in accordance with that policy.

Two other sets of cases deserve brief mention here. First, two domain name antitrust decisions have rejected the plaintiffs' claims for failure to define a proper economic market.[129] While these cases are not directly relevant to antitrust immunity, they do remind us of the important point that immunity is not all there is to antitrust law. Even if NSI is not immune from antitrust scrutiny, an antitrust plaintiff will have to prove all the elements of a section 2 claim in order to prevail. We discuss potential antitrust claims in more detail in Part III.

Second, two courts have considered whether NSI is a state actor in a different context: whether it must conform its conduct to the Constitution, and in particular to the First Amendment. In both cases, the court concluded that NSI was not a state actor for First Amendment purposes.[130] The courts emphasized the facts that registering domain names is not a traditional governmental function,[131] that the government did not impose restrictive regulatory oversight on NSI, and that the "nexus" between the government and NSI was not sufficiently close to find that the two were in a symbiotic relationship.[132] The standards for state action in the First Amendment context are different than in the antitrust context,[133] and the courts' conclusions are certainly contestable on their merits.[134] Still, it is interesting that in the constitutional context, courts have minimized the extent to which NSI acts at the government's behest. The facts they cite may turn out to be quite relevant to the antitrust immunity inquiry as well.

### C. Antitrust Immunity for ICANN?

Given this legal background, and what we learned about ICANN's relationship to the government in Part I, what are ICANN's prospects for antitrust immunity? As an initial matter, it seems safe to say that ICANN will not be able to rely on an absolute form of federal instrumentality immunity. Congress has not created an express exception to the antitrust laws for ICANN. Indeed, it has not spoken at all on the subject. So if ICANN is to be immune from antitrust suit, it must be because of its contracts with the Department of Commerce. But appellate courts so far have not endorsed the theory that any government contractor is entitled to absolute immunity. Rather, the most likely approach will be one akin to the state action doctrine: a case-by-case analysis of whether ICANN's actions were pursuant to a clearly articulated governmental policy to displace competition and were actively supervised by the government.

---

[129] Weber v. National Football League, 112 F. Supp. 2d 667 (N.D. Ohio 2000) (no relevant economic market for subset of domain names that constitute NFL trademarks); Smith v. Network Solutions, Inc., 135 F. Supp. 2d 1159 (N.D. Ala. 2001) (no relevant economic market for expired domain names).

[130] National A-1 Advertising, Inc. v. Network Solutions, Inc., 121 F. Supp. 2d 156 (D.N.H. 2000); Island Online v. Network Solutions, Inc., 119 F. Supp. 2d 289 (E.D.N.Y. 2000).

[131] The courts disagreed on this point. Compare *National*, 121 F. Supp. 2d at 166 (registering domain names is a traditional governmental function) with *Island Online*, 119 F. Supp. 2d at __ (holding the opposite). The D.C. Circuit weighed in on the latter side in *Thomas*, 176 F.3d at 511.

[132] *Id*.

[133] For a discussion of the First Amendment standards, see, e.g., Erwin Chemerinsky, *Rethinking State Action*, 80 **Nw. U. L. Rev.** 503 (1985); Paul Brest, *State Action and Liberal Theory*, 130 **U. Pa. L. Rev.** 1296 (1982).

[134] *See* Froomkin, *supra* note 5, at 113-25 (arguing that ICANN is a state actor for constitutional purposes); *see generally* Paul Schiff Berman, *Cyberspace and the State Action Doctrine: The Cultural Value of Applying Constitutional Norms to Private Regulation*, 71 **U. Colo. L. Rev.** 1263 (2000).

While the issue is not free from doubt, we are skeptical that all of ICANN's conduct can meet that test. ICANN does have a case on the clearly articulated government policy prong, but the facts currently in the public record suggest that it would have a very hard time showing the necessary degree of active government supervision and involvement in its implementation of that policy. Of course, it is always possible that ICANN would be able to demonstrate that the government has had a far greater behind-the-scenes involvement in ICANN's decisions than either DoC or ICANN has admitted; at present, however, we take the parties at their word that since its formation DoC has given ICANN very great independence.

The White Paper can be used to argue both sides of the "clearly articulated government policy" test. On the one hand, the White Paper itself considered and rejected the idea that 'NewCo,' as it then was, should be given antitrust immunity. Indeed, in the White Paper the government argued that "[a]pplicable antitrust law will provide accountability to and protection for the international Internet community. Legal challenges and lawsuits can be expected within the normal course of business for any enterprise and the new corporation should anticipate this reality."[135] This seems if anything a fairly clearly articulated policy that there *not* be antitrust immunity.[136]

On the other hand, the White Paper also contained a number of policy directions for NewCo, instructions that ICANN has on the whole faithfully followed. For example, in the White Paper the Department of Commerce clearly articulated a view that the DNS needed an anti-cybersquatting policy, and stated that the policy, whatever it was, should be put into place by a new non-profit corporation that took over administration of the DNS. That said, the White Paper had relatively little to say about the details.[137] Whether this general, but emphatic, statement in a legally non-binding 'policy statement', and the

---

[135] White Paper, *supra* note 22.

[136] But *see* Lebron v. National R.R. Passenger Corp., 513 U.S. 374 (1995) (holding that Congressional statement that otherwise public corporation was private did not make it private for First Amendment purposes).

[137] The White Paper stated:
"[T]he U.S. Government recommends that the new corporation adopt policies whereby:
  1) Domain registrants pay registration fees at the time of registration or renewal and agree to submit infringing domain names to the authority of a court of law in the jurisdiction in which the registry, registry database, registrar, or the "A" root servers are located.
  2) Domain name registrants would agree, at the time of registration or renewal, that in cases involving cyberpiracy or cybersquatting (as opposed to conflicts between legitimate competing rights holders), they would submit to and be bound by alternative dispute resolution systems identified by the new corporation for the purpose of resolving those conflicts. Registries and Registrars should be required to abide by decisions of the ADR system.
  3) Domain name registrants would agree, at the time of registration or renewal, to abide by processes adopted by the new corporation that exclude, either pro-actively or retroactively, certain famous trademarks from being used as domain names (in one or more TLDs) except by the designated trademark holder.

Department's subsequent praise for the UDRP, constitutes a sufficiently clear federal policy that there should be a UDRP certainly could be debated. Any such debate would be enriched by arguments that even if DoC had a policy favoring some kind of uniform dispute policy, it lacked the statutory authority to make such a policy. The *ultra vires* argument rests on the assertion that DoC has no specific statutory authority or obligation to control the DNS or the alpha root,[138] and the observation that the Administrative Dispute Resolution Act (ADRA) generally prohibits an agency from requiring arbitration to settle "the resolution of an issue in controversy that relates to an administrative program."[139] The problem here for ICANN is that if DoC directly required the UDRP it probably[140] violated ADRA. If it didn't directly require the UDRP, it's hard to argue that the federal policy was sufficiently explicit to meet the 'clear articulation' and 'active supervision' tests. Similarly, it is evident that DoC intended for one of ICANN's tasks to be the selection of new gTLDs, but other than defining general principles designed to guide ICANN, DoC did not instruct it as to how to go about picking TLDs or registries.

Even if ICANN were able to prevail on the "clearly articulated" prong, on the facts publicly available its chances of prevailing on the "active supervision" requirement look slim. Although DoC may have set out specific tasks for ICANN to achieve such as the prevention of cyberpiracy, both ICANN and DoC have asserted that ICANN acts independently of US Government control. Indeed, ICANN seems to be a paradigmatic case of a contractor left "free to perform the contract in any number of ways".[141] There are no defined procedures by which DOC reviews ICANN's work,[142] and even in the case of additions to the root, where DoC retains final authority to alter or countermand ICANN's decisions,[143]

---

4) Nothing in the domain name registration agreement or in the operation of the new corporation should limit the rights that can be asserted by a domain name registrant or trademark owner under national laws."
White Paper, *supra* note 22.

ICANN's ultimate plan substantially complied with 1 and 2, but not 3. It attempted to comply with 4, although how successful it was is hotly debated. *See* Froomkin, *supra* note 87.

[138] Such at least was the GAO's view *See* GAO Report, *supra* note 48.

[139] Administrative Dispute Resolution Act (ADRA), 5 U.S.C. §§ 571-83 (1994). An "issue in controversy" is defined as "an issue which is material to a decision concerning an administrative program of an agency, and with which there is disagreement" either "between an agency and persons who would be substantially affected by the decision" or "between persons who would be substantially affected by the decision." *Id.* §§ 571(8)(A)-(B). See Froomkin, *supra* note 5, at 135-36.

[140] There being no relevant caselaw, it remains possible for DoC to argue that by using a contractor to execute its policy it somehow took itself out of ADRA's reach.

[141] One of us has argued that ICANN is in fact *too* free of government control, and that DoC's grant of so much discretion to ICANN amounts to a violation of the non-delegation doctrine in Carter v. Carter Coal, 298 U.S. 238 (1936). *See* Froomkin, *supra* note 5.

[142] *See* Froomkin, *supra* note 5, at 107-13; Dinwoodie & Helfer, *supra* note 5 (arguing that ICANN is not subject to effective scrutiny).

[143] *See supra* note 53.

DoC has stated that it does not intend to exercise any substantive review.[144] DoC *has* intervened in ICANN policy making on rare occasions, but these primarily concerned ICANN's relationship with NSI/VeriSign, another government contractor. In 1999, DoC was intensively involved in brokering a deal between ICANN and NSI, in which NSI agreed to recognize ICANN's authority over it in exchange for an extension of its monopoly on the .com, .net and .org registries and certain limits on ICANN's freedom to regulate it.[145] Another intensive intervention came when several influential legislators objected to ICANN's proposed revisions to the ICANN-NSI/VeriSign contract, a change that required Commerce Department approval under the earlier set of agreements.[146] DoC and the US Justice Department stepped in and altered the agreement to reflect antitrust concerns arising from VeriSign's retention of ownership in both the dominant registries and the dominant registrar.[147]

In contrast, other than the statements in the White Paper, there is little in the public record to suggest that DoC instructed ICANN as either to the content of the UDRP or its procedural rules, or the ways in which ICANN should manage the selection of arbitration service providers. The main signs of continuing DoC involvement have been: (1) in July, 1999, a DoC official told a House Subcommittee that DoC had been consulting with ICANN before ICANN's major decisions, such as ICANN's proposal to charge a fee of $1 per domain name;[148] (2) in the June, 2000, MoU, DoC promised to devote more than a quarter of a million dollars in staff time and expenses to monitoring and helping ICANN, which DoC estimated would equal half-time dedication of four or five full-time employees;[149] and, in July, 2000, ICANN's Board passed resolution of thanks to outgoing NTIA official Becky Burr mentioning her "enormous contributions."[150] Other than these, there is little sign that the government has had a role in

---

[144] *See supra* note 54.

[145] *See* Froomkin, *supra*, note 5, at 89-91.

[146] *See* House Democrats Up the Ante on ICANN/VeriSign Deal, http://www.icannwatch.org/article.php?sid=154; Letter, http://www.house.gov/commerce_democrats/press/107ltr53.htm; US House Leaders Warn on VeriSign Deal, http://www.icannwatch.org/article.php?sid=72.

[147] *See* SPIN CYCLE: Commerce Release on Final ICANN/VeriSign Domain Name, http://www.icannwatch.org/article.php?sid=159 (quoting DoC press release).

[148] Letter from Andrew J. Pincus, DoC General Counsel, to Rep. Tom Bliley, Chairman, United States House Committee on Commerce (July 8, 1999), Information Administration, http://www.ntia.doc.gov/ntiahome/domainname/blileyrsp.htm.

[149] Memorandum of Understanding, *supra note* 41, at 7.

[150] "Along with her colleagues at the Department of Commerce, she played an essential facilitating role in not only the creation of ICANN, but also in its creation of contractual relationships with many of the important elements of the Internet community which have been and will be instrumental in its continued viability as an effective global, private sector, consensus creation body."

"It would not be an overstatement to conclude that, without the enormous contributions of Becky Burr, ICANN would not be here today, or at a minimum would not have made the very significant progress that is reflected at this meeting. She could not have done it alone, but we could not have done what we have done without her tireless devotion to the objective of a viable and effective ICANN."
Preliminary Report: Meeting of the ICANN Board in Yokohama, Res. 00.59-00.62, at http://icann.org/minutes/prelim-report-16jul00.htm (July 16, 2000) at 12-13.

supervising the (increasingly criticized[151]) administration of the UDRP since its adoption, other than general cheerleading,[152] although one presumes those DoC employees were doing something.

Similarly, there is nothing in the public record to suggest that DoC took an active role in ICANN's selection of new gTLDs. DoC's role appears to have been limited to giving ICANN authority to select new gTLDS in its initial contracts, and in its rubber-stamp approval of ICANN's choices, with little or nothing in between. The high water mark of DoC's intervention appears to have been a recent letter from Secretary of Commerce Evans to ICANN -- long after the first round selection process was over -- to urge it to approve the next round of domains more quickly.[153] The facts that Secretary Evans felt a need to write to ICANN, rather than just instruct it, and that the letter appears to have had no effect whatsoever, both argue strongly that ICANN's selection of new gTLDs is not subject to close supervision by DoC.

ICANN's resistance to alternate roots follows the same pattern. Again, the policy arguably has its origins in the White Paper, which stated that "The introduction of a new management system should not disrupt current operations or create competing root Systems."[154] Similar language did not, however, get included in the ICANN-DoC MoU.[155] In particular, there is no evidence that ICANN's refusal to even consider applications from firms that enabled alternate roots was required by the government.[156] This is particularly important because ICANN's revenue base depends on its being in charge of the only root of any importance. ICANN thus stands to gain from keeping its monopoly, and the government effectively delegated market control to a private party with an interest in the outcome. While the government can do this, the delegee is entitled to antitrust immunity only if the government actively supervises its conduct. Where private conduct directly restricts competition, that supervision must include direct control over the price or output setting, not merely a generalized delegation.[157] This does not appear to be the case with ICANN.

---

[151] *See*, *e.g.*, Froomkin, *supra* note 87; Michael Geist, *Fair.com?: An Examination of the Allegations of Systemic Unfairness in the ICANN UDRP*, http://aix1.uottawa.ca/~geist/geistudrp.pdf (August 2001); Milton Mueller, *Rough Justice: A statistical assessment of ICANN's Uniform Dispute Resolution Policy*, http://dcc.syr.edu/roughjustice.htm; Elizabeth G. Thornburg, *Going Private: Technology, Due Process, and Internet Dispute Resolution*, 34 **U.C. Davis L. Rev.** 151 (2000).

[152] An NTIA report last year praised the UDRP as "an efficient, inexpensive procedure for the resolution of disputes." U.S. Dept. of Commerce, National Telecommunications and Information Administration Annual Report 2000, http://www.ntia.doc.gov/ntiahome/annualrpt/2001/2000annrpt.htm.

[153] *See* David McGuire, Newsbytes, *Commerce Department Urges ICANN To Add More New Domains*, May 25, 2001, http://www.newsbytes.com/news/01/166139.html.

[154] White Paper, *supra* note 22, at 31,743.

[155] Compare the White Paper with the MOU.
"The U.S. Government should end its role in the Internet number and name address system in a manner that ensures the stability of the Internet. The introduction of a new management system should not disrupt current operations or create competing root systems. During the transition and thereafter, the stability of the Internet should be the first priority of any DNS management system. Security and reliability of the DNS are important aspects of stability, and as a new DNS management system is introduced, a comprehensive security strategy should be developed."
White Paper, *supra* note 22, at 31,743.
"This Agreement promotes the stability of the Internet and allows the Parties to plan for a deliberate move from the existing structure to a private-sector structure without disruption to the functioning of the DNS. The Agreement calls for the design, development, and testing of a new management system that will not harm current functional operations."
MOU, *supra* note 41, at § C.1

[156] For a discussion of alternate roots and their competitive implications, *see infra* __.

It is true that NSI has so far avoided antitrust liability for its actions in running the DNS during a prior era. But ICANN may not fare so well. Most of the cases against NSI were in fact ultimately resolved on the antitrust merits, not on grounds of antitrust immunity.[158] The one case ultimately finding immunity relied on the fact that NSI was specifically directed to engage in the challenged practice by the government. ICANN may be able to point to similar government mandates in a few cases, but surely cannot justify all its policies in this way. As one court put it, "the government's role in the Internet is deliberately waning. By design, the private sector is assuming an ever-increasing role in determining relevant policies and protocols, and domain name registration is now a competitive endeavor . . . ."[159] ICANN was intended to get the U.S. government out of the business of running the DNS. While the government certainly hasn't succeeded completely in disentangling itself from the domain name system, it gives less policy direction and less direct oversight to ICANN than it did to NSI in the mid-1990s. With that increased authority comes responsibility under the antitrust laws. ICANN's actions may or may not violate the antitrust laws; we turn to that subject in the next Part. But it is not entitled to ignore those laws altogether.[160]

## III. Is ICANN Violating the Antitrust Laws?

Assuming that ICANN is not immune from antitrust liability,[161] the next step is to consider the merits of antitrust claims likely to be brought against it.[162] In this part, we consider the merits of three likely antitrust challenges: a claim that the DNS and/or the TLDs are an essential facility to which ICANN must open access on reasonable and nondiscriminatory terms; a claim that ICANN's refusal to accredit registrars who are affiliated with alternative or competitive roots is illegal; and a claim that ICANN's insistence on registrars' adherence to uniform mandatory dispute resolution policies is illegal.

---

[157] *See, e.g.,* A.D. Bedell Wholesale Co. v. Philip Morris, __ F.3d __, 2001-1 Trade Cases ¶ 73,310 (3d Cir. 2001) (state action doctrine doesn't provide immunity from allegation that state government-tobacco settlement facilitates a cartel; while the government clearly articulated its policy of restricting production of cigarettes, the states "lack oversight or authority over the tobacco manufacturers' price and production levels. These decisions are left entirely to the state actors." This constituted inadequate state supervision).

[158] *See supra* notes __-__ and accompanying text.

[159] *National A-1 Advertising*, 121 F. Supp. 2d at 166. *See also* Zittrain, *supra* note 7, at 1092 (describing ICANN as taking a "middle path" between public and private status).

[160] The decreasing role of the government also makes it likely that VeriSign, NSI's successor company, will no longer enjoy immunity for its conduct. Indeed, the May 2001 revision of the contract between ICANN and VeriSign makes it clear that the Commerce Department does not intend to immunize VeriSign from antitrust scrutiny. *See, e.g., Department of Commerce Approves ICANN Registry Agreements with VeriSign, Inc.*, 6 **Elec. Comm. & L. Rpt.** 567 (2001).

[161] As we noted in part II, there is some question as to ICANN's immunity. But for the remainder of this part and the next, we will assume that ICANN is not categorically immune from antitrust liability.

### A. Principles of Antitrust Law

Antitrust law treats unilateral conduct under the law of monopolization. The governing statute is section 2 of the Sherman Act, which sweeps broadly to condemn "every person who shall monopolize, or attempt to monopolize," a relevant market.[163] The concept of monopolization embodies two crucial principles. First, to be liable under section 2 of the Sherman Act, a defendant must be a monopolist, or at least be likely to become a monopolist. Antitrust law does not generally scrutinize the unilateral conduct of individuals or companies; those who hold a monopoly position in a market are an exception to this general rule. Second, the mere possession or even acquisition of a monopoly alone is not illegal.[164] Rather, the offense of monopolization requires not just a monopoly, but some sort of anticompetitive conduct designed to acquire or maintain that monopoly.

Thus, in *United States v. Grinnell Corp.*,[165] the Court defined monopolization under section 2 as follows:

> (1) the possession of monopoly power in the relevant market, and
>
> (2) the willful acquisition or maintenance of that power as distinguished from growth or development as a consequence of a superior product, business acumen, or historic accident.[166]

These limitations are designed to balance two competing policy interests. On the one hand, the antitrust law reflects an economic judgment that competition is desirable and monopolies are undesirable. Competition is good for a variety of reasons. Basic economics teaches that firms in competition will produce more and price lower than monopolists. Monopolists not only take money away from consumers by raising prices, but they impose a "deadweight loss" on society by reducing their output below the level which consumers would be willing to purchase at a competitive price.  As a result, some transactions that would make economic sense (because consumers value the product at more than it would cost to produce it) do not occur.[167]  Monopoly has other problems as well.  It inherently reduces consumer choice, and monopolists have fewer incentives to innovate than do competitive firms.

---

[162] While we are aware of no reported antitrust decisions involving ICANN to date, the proliferation of claims against NSI – including some still being filed after ICANN took over, see Chrysalis Vineyards v. U.S. Department of Commerce, No. 00-1330-A (E.D. Va. filed Sept. 7, 2000) – suggest that such cases are coming.

[163] 15 U.S.C. §2.

[164] *See, e.g.,* United States v. Aluminum Co. of America, 148 F.2d 416 (2d Cir. 1945) (acquisition of monopoly power by "superior skill, superior products, natural advantages, economic or technological efficiency," and other legal means is not illegal).

[165] 384 U.S. 563 (1966).

[166] *Id*. at 570-71.

At the same time, the law cannot merely forbid all monopolies.  Some monopolies result from natural economic conditions that permit only one firm to operate efficiently in a given market.  Such "natural monopolies" include local distribution of electric power or telephone service.  Other monopolies result from vigorous competition on the merits -- precisely the sort of behavior the antitrust law is designed to encourage.  Still other monopolies result from a firm's innovation, either because the innovation makes the firm more efficient or because legally granted intellectual property rights give the firm a certain measure of control over a market.[168]  If the law forbade every monopoly, it would discourage innovation and competition by causing large companies to worry too much about questionable behavior.[169]  The section 2 cases attempt to strike a balance by prohibiting only monopolies acquired or maintained by anticompetitive means.

Section 1 of the Sherman Act forbids agreements in restraint of trade.  Courts have identified two basic types of agreements that may be in restraint of trade -- agreements among competitors (called "horizontal restraints") and agreements between buyers and sellers (called "vertical restraints").[170]  Vertical restraints are generally less threatening to competition than horizontal restraints.  With the exception of vertical price fixing, they are generally judged under the "rule of reason."  Under the rule of reason, courts balance the anticompetitive harms of a restraint against its procompetitive benefits.  Only those restraints which produce harms significantly in excess of benefits to competition are deemed unreasonable.

Horizontal restraints are more troubling because they may allow the participants to create a cartel, which can then behave anticompetitively, much as a monopolist would.  At first, most agreements between competitors were deemed illegal "per se," without any necessity for a weighing of harms and benefits to competition.  Today, the Supreme Court has retreated from that position, recognizing that certain agreements among competitors may be efficient and procompetitive.  Most horizontal restraints are now judged under the rule of reason.  Only certain forms of "naked" agreements to fix prices or divide territories remain illegal per se.  Nonetheless, it is fair to say that antitrust treats agreements among competitors more harshly than is unilateral conduct.

---

[167] For a detailed discussion of the economic intuition here, *see* 2A **Areeda & Hovenkamp, Antitrust Law** ¶¶ 402-415**.**

[168] We do not mean to suggest that intellectual property rights normally confer market power; far from it.  Normally they do not.  *See* I **Herbert Hovenkamp et al, IP and Antitrust** § 4.2 (forthcoming October 2001).  But intellectual property rights do sometimes provide competitive advantages to their owners.  When they do, it would undermine the intellectual property laws to make it illegal to take advantage of those rights.

[169] This is particularly true since, as we shall see, the remedies for an antitrust violation can include treble damages, structural relief breaking up a company, and even criminal penalties.

[170] Actually, the term "vertical restraints" refers to a whole class of transactions between companies in a vertical relationship in the chain of distribution, including dealers, franchisors, distributors, resellers, etc.

While the antitrust laws apply to acts "in commerce," it is clear that ICANN's non-profit status will not protect it. Antitrust law reaches non-profit concerns so long as they engage in activities that affect commerce;[171] ICANN clearly does so. Thus, we turn in the following sections to the substantive antitrust issues that are likely to be raised by ICANN's conduct to date.

### B. DNS as an Essential Facility

One sort of monopolization case departs from the general rules articulated above because it does not involve "conduct" at all in the affirmative sense. Courts sometimes hold that a monopolist has a duty to deal with competitors, or at least to continue a relationship once it has begun.[172] Under this doctrine, the monopoly owner of an "essential facility" for competition may be forced to give access to that facility to competitors on reasonable and nondiscriminatory terms.[173] The essential facilities doctrine is unique in that a monopolist's status (as the owner of the facility and a competitor in the market that relies on the facility) rather than any affirmative conduct determines liability.[174]

The essential facilities doctrine grew out of a number of cases in which one company (or a group of them) had exclusive control over some facility and used that control to gain an advantage over competitors in an adjacent or downstream market. Most of the canonical cases have this basic structure. Thus, in *Terminal Railroad*, a group of railroads jointly owned a key bridge over the Mississippi River and accompanying rail yard, and refused to give competing railroads use of the facilities.[175] In *Otter Tail*, the public utility that owned all the transmission lines into a municipality refused to allow the municipality to "wheel" power over those lines from outside plants, because the utility itself wanted to provide power to the municipality.[176] And in *MCI v. AT&T*, the pre-breakup Bell System refused to permit MCI to connect its long distance calls to the Bell System's local phone exchanges.[177] In each of these cases, the defendant owned a facility that could not plausibly be duplicated, and also participated in a competitive downstream market that required access to the facility. By denying access to the facility, the defendant either eliminated its downstream competitors or imposed significant costs on them.[178]

---

[171] *See, e.g.,* National Collegiate Athletic Ass'n v. Board of Regents, 468 U.S. 85, 100 n.22 (1984); _ **Areeda & Hovenkamp, Antitrust Law** ¶ __; Tomas J. Philipson & Richard A. Posner, *Antitrust and the Not-For-Profit Sector* (NBER working paper 8126).

[172] On the latter concept, *see* Aspen Skiing Co. v. Aspen Highlands Skiing Corp., 472 U.S. 585 (1985) (finding a refusal to continue dealing by a monopolist illegal in the absence of a legitimate business justification).

[173] For a detailed discussion of the essential facilities doctrine, *see* 3A **Areeda & Hovenkamp, Antitrust Law** ¶¶ 770-774 (2d ed.).

[174] The monopolist in an essential facilities case may be thought to have "acted" in some sense, by refusing to deal or to continue dealing with a competitor. But generally speaking a unilateral refusal to deal is not the sort of anticompetitive conduct that the antitrust law is concerned with.

[175] United States v. Terminal R.R. Assn., 224 U.S. 383 (1912).

[176] Otter Tail Power Co. v. United States, 410 U.S. 366 (1973).

[177] MCI Comm. Corp. v. AT&T, 708 F.2d 1081 (7th Cir. 1983).

In *MCI*, the court set out a four-part test for an essential facilities claim:

(1) control of the essential facility by a monopolist; (2) a competitor's inability practically or reasonably to duplicate the essential facility; (3) the denial of the use of the facility to a competitor; and (4) the feasibility of providing the facility.[179]

If such a claim is made out, the defendant will be obligated to provide access to the facility on reasonable and nondiscriminatory terms.

Under this test, the defendant must be a monopolist, and the facility must be "essential" in the sense that the competitor needs access to it in order to compete. An essential facility will therefore normally be an input into the competitive market – some component that must be used in providing the competitive product or service. The need must be substantial, however – inconvenience or cost increase resulting from unavailability should not suffice.[180] The court's test also offers a defense of legitimate business justification, by permitting the defendant to show that it wasn't feasible to provide access to the facility. The "reasonable and nondiscriminatory terms" language also limits the defendant's obligation in circumstances where particular plaintiffs cannot afford to pay, aren't willing to pay a reasonable price, or the like.[181]

While the *MCI* court doesn't discuss it directly, it seems important to add that withholding an essential facility is illegal only if it has the effect of foreclosing competition in the downstream market, and therefore of helping the defendant to acquire or maintain a monopoly in that market. Thus, the owner of the facility in question must be vertically integrated into the market in which competition is being foreclosed. *Otter Tail* and *MCI* both had such a characteristic.[182] In the absence of such a market effect, condemning a truly unilateral refusal to deal could open the door to all sorts of claims in which competition is not really at stake.

---

[178] A very different sort of essential facility-type claim is envisioned by those few cases that impose a duty to continue dealing. For example, in *Aspen Skiing Co. v. Aspen Highlands Skiing Corp.*, 472 U.S. 585 (1985), the Court held that a ski company that owned 3 of the 4 mountains in a local area was obligated to continue offering a multi-area skiing pass with its sole competitor in that local area. While the Court did not discuss the case in essential facilities terms, there is no other antitrust concept that readily fits these circumstances. By avoiding the use of essential facilities language, however, the Court short-circuited inquiry into how important the multi-area pass actually was to competition.

[179] *MCI*, 708 F.2d at 1132-33.

[180] *See* Alaska Airlines v. United Airlines, 948 F.2d 536, 544-46 (9th Cir. 1991) (airline computer reservation system was not an essential facility because airlines could compete without it, albeit at higher cost).

[181] Whether this defense would extend to other sorts of business justifications for refusing to deal is not clear.

[182] It is not clear that *Terminal Railroad* fits easily in this framework, but that case may be complicated by its reliance on a conspiracy between different railroads in violation of section 1.

The essential facilities doctrine has come in for serious criticism. Many prominent antitrust scholars have argued that the doctrine should be abolished outright.[183] Others who favor the continued existence of the doctrine nonetheless concede that it is properly applied only in rare cases.[184]

Is the legacy-root server an essential facility to which ICANN must provide access? One court addressed this issue in a suit against NSI, though its decision leaves a number of issues unresolved. In *Thomas v. Network Solutions*,[185] the plaintiffs were entities who had registered domain names through NSI. The D.C. Circuit held against them on the essential facilities claim, not because access to the root server wasn't essential (a question it did not decide), but because the plaintiffs in that case did not compete with NSI in a downstream market and so could not demonstrate a required element of an essential facilities claim.[186]

A more likely essential facilities claim is one brought by a competitor. Under early pre-ICANN market structure, such a claim was fairly easy to envision. The government controlled the root zone file that was relied on by the alpha root server and all downstream copies of it. NSI controlled both the authoritative .com registry and was the exclusive registrar for .com and the other open gTLDs. Entry in the root zone file allowed firms to be registries; NSI's entry for .com meant that an entry in its database was necessary to permit others to access your Web site by typing an alphanumeric URL; if your domain name was not in the registry whose address was found on the master list, no one relying on DNS servers in the legacy root could find you by entering that domain name.[187] If a plaintiff sought to compete as a registrar by taking registrations for .com, it would be stymied by NSI's refusal to enter such competing registrations in the only authoritative registry for .com listed in the root zone file. Thus, NSI would have denied access to a facility it controlled (the .com registry), essential to competition, to which it could feasibly have provided access, with the effect of perpetuating its dominance in a separate product market – the market for registration services.

Over time, NSI was forced to allow other registrars to sell registrations. Even then, there were allegations that NSI's registry gave various sorts of preferential access to its registrar, and that the NSI registry's price (set in a contract with the US Government) exceeded fair market value. These problems stemmed from NSI's vertical integration. NSI controlled the .com registry, and also competed with other registrars in the market for registration services. It thus had the classic structure for an essential facilities claim. The obvious solution was to separate control of the registry from control of the registrar. Indeed, getting an agreement that NSI would divest itself of either the registrar or the registry by May 10, 2001 was supposed to be one of DoC and ICANN's major achievements, although when the time came and NSI threatened to divest itself of the registrar, and then affiliate with another one, ICANN blinked. It instead accepted VeriSign/NSI's proposal to divest itself of .org and .net within a few years while keeping .com. The government then further modified the agreement to require auditing of the 'firewall' between the registry and the registrar and a few other antitrust-inspired changes.[188]

---

[183] *See, e.g.,* Philip Areeda, *Essential Facilities: An Epithet in Need of Limiting Principles*, 58 **Antitrust L.J.** 841 (1989); **Herbert Hovenkamp**, **Federal Antitrust Policy** at §7.7 (2d ed. 1999) ("The so-called essential facility doctrine is one of the most troublesome, incoherent and unmanageable of bases for Sherman § 2 liability. The antitrust world would almost certainly be a better place if it were jettisoned . . . "); David McGowan, *Regulating Competition in the Information Age: Computer Software as an Essential Facility Under the Sherman Act*, 18 **Hastings Comm/Ent. L.J.** 771 (1996).

[184] *See* Mark A. Lemley, *Antitrust and the Internet Standardization Problem*, 28 **Conn. L. Rev.** 1041, 1085-86 (1996).

[185] 176 F.3d 500, 509 (D.C. Cir. 1999).

[186] *Id*. *Cf.* America Online v. Greatdeals.net, 49 F. Supp. 2d 851, 862 (E.D. Va. 1999) (complaint properly alleged that email access to AOL subscribers was an essential facility, but was nonetheless dismissed because the plaintiff and defendant did not compete).

[187] They could, however, reach your web site with a browser by typing in the IP number. E-mail works slightly differently, and there are some e-mail programs that simply cannot send e-mail to an address at an IP number, but these are relatively rare. Although it is not part of the minimum specifications, most e-mail programs can send mail to an address of the form user@[129.171.97.1].

ICANN's control of access to the root raises different issues from NSI's because ICANN acts as neither a registrar or a registry.[189] ICANN's has power over registries and (through them) registrars; its control is greatest over those seeking ICANN's approval to enter the legacy root. ICANN has at least one direct financial incentive to limit the number of gTLDs to the root. So long as there is a shortage of gTLDs, firms will pay ICANN substantial sums simply to be allowed to apply for consideration. Indeed, last year ICANN was able to require applicants to pay it a non-refundable $50,000 fee, "intended to cover ICANNs costs of receiving and evaluating the application, including performing technical, financial, business, and legal analyses, as well as ICANNs investigation of all circumstances surrounding the applications and follow-up items."[190] Forty-seven applicants purchased what amounted to expensive lottery tickets, but ICANN selected only seven to receive a TLD. The greater the number selected, the less ICANN would have the leverage to require a similar payment from the next round of applicants.

Other than the premium it can demand from would-be entrants, however, it is debatable whether ICANN itself, as opposed to incumbent registries, has a financial incentive to limit the number of new entrants to the root. ICANN annually sets its financial needs, and assesses income from registries and registrars, who pay according to various formulas that in part reflect their market shares. All other things being equal,[191] ICANN may have a financial incentive to increase the number of registrants, since that spreads the costs and increases the amounts it can levy without occasioning protest, which should argue for more TLDs since these should tend to increase total registrations. To the extent that new TLDs just shift a constant number of registrations around registries, ICANN should be neutral, unless the very small number of registries allows them to charge a premium price, and lets ICANN demand part of that rent for itself. Only if new TLDs were to so significantly increase supply in the relevant market that it substantially depressed the prices charged by registries or registrars might ICANN's income stream be affected.[192] Because ICANN does not compete directly in either the registry or registrar markets, it does not conform to the classical structure of an essential facilities case.

---

[188] *See supra* text at notes --.

[189] Actually, ICANN has hosted the registries for new gTLDs as they come on stream. The reasons for this are not clear. Some have suggested it's a subsidy to the registries; others that ICANN wants an excuse to buy computers, or to learn how to run a registry in case it ever needs to rescue a failing one. ICANN's announcement was short on details, but describes it as a temporary measure.

[190] *See* ICANN, New TLD Application Process Overview, http://www.icann.org/tlds/application-process-03aug00.htm (Aug. 3, 2000) (noting "The application fee is non-refundable and ICANN's only obligation upon accepting the application and fee is to consider the application.").

[191] One way in which they are not equal is the administrative cost of dunning large numbers of registrars. ICANN has been moving to a funding strategy that concentrates on registries since there are fewer of them.

ICANN's relationship with the alternate root operators presents a more complex issue in market definition and definition of market participants. On the one hand, the alternate root operators as a group are ICANN's true competitors in that they create opportunities (currently, tiny ones) for new TLDs and new registries. Plus, entrants in their roots do not pay ICANN's levies, giving ICANN a financial motive to fear growth in their market penetration -- a growth that network effects suggest would be likely to take off once it reached some distant critical mass.[193] On the other hand, it is a little difficult to identify any individual person or firm as ICANN's competitor. The alternate root operators work in a far more decentralized fashion than ICANN. There is no central policymaking body for the alternate roots as a group, and most of the major groups have only loose coordination bodies seeking to head off name collisions. So while the alternate root based registries as a group are ICANN's competition, individually the operators are more like competition for ICANN-approved registries than for ICANN itself.

The most important competitive relationship between ICANN and the alternate root operators arises from the competition for the name space. Traditionally, alternate root operators have worked with considerable (but not total) success to avoid creating "name collisions"[194] – two TLDs that use same character string. They have also avoided creating TLDs that conflict with legacy root, since users of the alternate root namespace also use the legacy root. Because ICANN does not recognize the existence of the alternate roots, however, it has no compunction about approving TLDs that use a string already in use in an alternate root. Thus, for example, ICANN accepted and debated applications for the .web TLD from parties other than IODesign, which has been running it as an alternate root since 1996,[195] although it chose not to assign that name to anyone. ICANN did, however, assign the .biz string to NeuLevel when there was already a very small functioning alternate root by that name.[196] In the view of the original .biz's partisans that makes ICANN the "name collider"; whoever is right, both names cannot be in the same root, forcing alternate root operators to choose, for the first time, whether to abandon one of their own or to offer data conflicting with that used by the legacy root rather than simply supplemental to it. Whatever the rights and wrongs of all this, it demonstrates that ICANN has a strong competitive effect on alternate TLDs with which its new entries to the root collide.

---

[192] There might also be administrative costs to ICANN associated with new TLDs. ICANN has clearly found it difficult to negotiate contracts with the new TLDs, some of which have to date taken nine months more than originally envisioned; large numbers of TLDs might have staffing implications.

[193] *See generally* Michael Katz & Carl Shapiro, *Network Externalities, Competition, and Compatibility*, 75 **Am. Econ. Rev.** 424 (1985); Mark A. Lemley & David McGowan, *Legal Implications of Network Economic Effects*, 86 **Calif. L. Rev.** 479 (1998).

[194] *See infra* text at note .

[195] *See* Image Online Design, FAQ, https://www.webtld.com/info_faq.asp .

[196] *See* ICANNWatch.org, *.biz Is Tiny -- or Is It* (June 19, 2001), for competing accounts of the size and viability of the 'alternate' and pre-existing .biz.

How would an essential facilities claim fare under this new structure? It seems clear that ICANN controls access to the system by which the overwhelming majority of registrants obtain domain names. At the same time, the existence of alternate roots may make it less likely that ICANN actually controls access to a facility essential to competition. If the barriers to duplicating -- or more likely, supplementing -- the root are not extraordinary, an essential facilities claim will founder on the first and second elements. Network effects may make it more difficult to set up a competing root that draws many customers. But they will not make it impossible.[197] Further, even if the legacy root is essential at the root level, ICANN is not vertically integrated. Thus, even if ICANN's actions are consistent with a view that it seeks to punish anyone that tries to compete with it by running an alternate root, ICANN lacks the clear self-dealing incentive present in cases like *MCI* and *Otter Tail* because ICANN gets no direct financial benefit from choosing one registry over another.

A second problem with the essential facilities theory concerns the feasibility of providing access to the facility to everyone on reasonable and nondiscriminatory terms. If ICANN's concerns about the stability of a DNS with multiple gTLDs have any basis,[198] an essential facilities claim will founder on this fourth element. Even if those concerns are overstated, courts may well decide to defer to ICANN's expertise on the issue rather than take a chance with the stability of the domain name system. As a result, even if ICANN is determined to control access to an essential facility, we are skeptical that courts will require it to open that facility to all comers because of concerns that doing so would be impractical.

In short, it is unlikely that the legacy root will be determined an essential facility to which ICANN must provide access on reasonable and nondiscriminatory terms. This doesn't mean that ICANN will avoid liability for specific anticompetitive conduct, however; we discuss those issues in the sections that follow.

### C. ICANN's Decision to Limit gTLDs and Restrict Registries

---

[197] For an early discussion, *see* Lemley & McGowan, *Network Effects*, at 555 n.323:
> Unlikely is not impossible, however. There are some reasons to believe that even here, network effects might not prevent effective competition between standards. First, the switch to a new domain name system need not be a complex one. NSI cannot claim to own the basic protocols that govern the Internet. It might be relatively straightforward, therefore, for a concerted group of large Internet users to switch their allegiance in a public way, causing others to follow suit. Second, and more important, it might be possible to run a new DNS system *alongside* the existing one, so that a company could be on both systems at once. If this is feasible, IBM could be accessed through ibm.com via NSI, and through a different (*or conceivably even the same*) domain name on a different system. Which system a user used would depend on how she accessed the Net. Lock-in concerns are significantly alleviated to the extent that users can simultaneously use more than one standard, as we have seen.

[198] We discuss this issue in detail in the next section.

ICANN's decision to limit the number of new gTLDs created an artificial scarcity of domain names. It also limited the number of companies who could be registries, since the DNS as we know it assumes that there will be only one registry for each gTLD. ICANN's method of choosing registries presents rather serious antitrust issues.[199] ICANN's application document for would-be new gTLD registries,[200] and especially the accompanying "Criteria for Assessing TLD Proposals,"[201] made it clear that parties who dealt with ICANN's competitors would be rejected -- even though those competitors' market share was almost trivial.[202] ICANN warned applicants that they must "demonstrate specific and well-thought-out plans, backed by ample, firmly committed resources, to operate in a manner that preserves the Internet's continuing stability. The introduction of the proposed TLD should not disrupt current operations, nor should it create alternate root systems, which threaten the existence of a globally unique public name space."[203] ICANN's demand that all applicants for approval as a new gTLD registry first forswear 'alternate roots' is an exclusive dealing requirement.

Exclusive dealing arrangements are suspect under the antitrust laws because when entered into by a firm with a significant share of the market, they may foreclose options to competitors, driving them from the market entirely or raising their costs.[204] For example, a dominant manufacturer may be able to "lock up" a large number of retail outlets by demanding that those outlets deal exclusively to it. As a result, competing manufacturers may find it difficult or impossible to place their goods in retail stores.

Not all or even most exclusive deals are anticompetitive, however. Exclusive dealing arrangements can also serve the useful purpose of guaranteeing a manufacturer an ongoing source of supply or a continuing outlet for distribution. This in turn permits the manufacturer to make investments on a long-term basis. It may also facilitate quality control and monitoring of sales outlets by the manufacturer.[205] As a result, exclusive dealing arrangements are judged under the rule of reason.[206] They are illegal only if the firm insisting on the agreement has a sufficient share of the market that the agreement will foreclose a significant amount of competition. Even then, the agreement can be justified if the defendant can show procompetitive benefits that outweigh any foreclosure.[207]

---

[199] For a related concern, that competition among gTLDs is still minimal, *see* Kesan & Shah, *supra* note __, at 198-200.

[200] http://www.icann.org/tlds/new-tld-application-instructions-15aug00.htm

[201] http://www.icann.org/tlds/tld-criteria-15aug00.htm

[202] http://www.icann.org/icp/icp-3-background/response-to-new.net-09jul01.htm. *See also ICANN Chief Issues Statement Criticizing Existence of Alternative Domain Name Roots*, 6 **Elec. Comm. & L. Rpt.** 587 (2001).

[203] *Id*.

[204] For a general discussion of the competitive risks of exclusive dealing arrangements, *see* XI **Areeda & Hovenkamp** ¶ 1801.

[205] *See id.* ¶ 1802.

[206] Exclusive dealing arrangements in goods are governed by section 3 of the Clayton Act, 15 U.S.C. § 14. Because registries provide services, not goods, the relevant law is provided by section 1 of the Sherman Act, 15 U.S.C. § 1. But the legal standards are largely the same in any event. *See generally* XI **Philip Areeda & Herbert Hovenkamp, Antitrust Law** ¶ 1800c4-c5.

Whether ICANN's exclusive dealing requirement is legal depends on ICANN's market power and on whether the exclusive dealing requirement is on balance pro- or anticompetitive. ICANN unquestionably has control over the legacy root. Virtually all gTLDs (measured by use) are under ICANN's effective control, and ICANN's control over access to the legacy-root server creates rather substantial barriers to entry to alternate roots. Estimates of how many people use alternate roots vary, but the numbers of people using true alternate name resolution services is probably well under 1% of all Internet users, and the number of domain name registrants in the true alternate roots was very small indeed. Even today, what is presumed to be the largest 'alternate' registry, .web, boasts only about 26,000 registrations.[208] The small take-up is hardly surprising, given that the alternate roots suffer from a classic network effect,[209] and that a registration in an alternate root is of relatively little value in the absence of a critical mass of fellow users able to resolve the name. The existence of this network effect, coupled with ICANN's control over the dominant root, makes ICANN's exclusive deal particularly effective. By denying alternate roots the right to participate in running gTLDs in the legacy root, ICANN keeps those alternate roots marginalized, and makes it far less likely that they will ever achieve that critical mass. The foreclosure in question here is not substantial in percentage terms simply because ICANN's control is so complete.[210] But it effectively forecloses the most likely source of competition for ICANN's legacy root.

Understanding the competitive significance of ICANN's requirement that applicants demonstrate that their systems would not be compatible with "alternate root systems" requires a short detour into DNS architecture. Recall that ICANN gets its power from its control over a key Internet chokepoint -- the content of the legacy root file. Users can try to avoid the effects of this chokepoint by using so-called "alternate roots": rather than getting their name resolution service from a member of the legacy root hierarchy they instead get DNS service from someone else who gets her data from a different root file with more or different entries.

At the time ICANN was choosing among the applicants for new gTLDs, the most commonly deployed type of alternate root were super-sets of the legacy root.[211] These alternate DNS services, such as the ORSC, either cache the legacy root's data or route queries to it when users seek to resolve a domain name in, say, .com.  But where legacy root servers would give an error message for lighting.faq, or law.web, these services send the queries to private registries that operate without the Department of Commerce's imprimatur.  Most of the alternate roots in operation belong to a loose cooperative network that works on first-come-first-served principles; this cooperative encourages peering and minimizes, but does not entirely eliminate,[212] the problem of "colliders" -- two or more registries claiming to be the authoritative source of registrations in a particular TLD.[213]

---

[207] *See, e.g.,* Tampa Electric Co. v. Nashville Coal Co., 365 U.S. 320, 334 (1961); XI **Areeda & Hovenkamp** ¶ 1801i. For criticism of the balancing approach, and suggested alternatives, see *id*. at ¶ 1822b.

[208] http://www.icannwatch.org/article.php?sid=212 .

[209] On the role of network effects in cementing the control of the legacy root, *see* Lemley & McGowan, *supra* note __, at 553-55.

[210] The absence of a significant foreclosure percentage often dooms exclusive dealing claims. *See* XI **Areeda & Hovenkamp** ¶ 1821d1. But the defendant rarely has the almost complete control over the market that ICANN has in this case. We do not think that ICANN's success in dominating the market renders its reliance on exclusive dealing arrangements less problematic.

[211] A subsequent entrant to the non-ICANN domain name market actually offers a service that is a complex blend of legacy and alternate root services. *See infra* text at (discussing New.net).

ICANN's action to foreclose any chance that new gTLD registries would in any way assist the alternate roots obviously benefited ICANN itself.  ICANN's authority and revenues flow from its contracts with registrars and registries.  By ensuring that the alternate roots remain shut out from the biggest players, ICANN exacerbates the network effects that keep its primary competition small.[214]  Registries in the ICANN system who would not be interoperating with alternate roots also benefit, since they are protected from that competition.   On the other hand, registrars and customers lose out, since there are fewer domain names to sell, and less competition between registries.  The exclusive dealing requirement therefore seems anticompetitive on its face.

Exclusive dealing arrangements entered into by monopolists that foreclose significant competition are presumptively illegal.  Even so, the defendant will escape antitrust liability if it can demonstrate a legitimate procompetitive justification for its behavior.  In this case, whether ICANN can make such a showing will depend on the technical merits of its argument that alternate roots create instability.[215]

The case for ICANN's technical rationalization of its policy against alternate roots relies on the weight of establishment technical opinion, especially the opinion of the influential Internet Architecture Board (IAB).  There is, however, a case to be made that the rationale is pretextual.  Proponents of alternate roots certainly disagree with it; and ICANN's own protocol standards body recently refused to endorse the IAB opinion.  ICANN's response towards its first vigorous competitor, new.net, has been very aggressive.

The technical case against alternate roots rests in large part on the belief that domain names should always resolve to the same resource[216] regardless of who is accessing it and where they are located.  If competing roots have name collisions for a TLD, i.e. if there is more than one registry for a given TLD taking competing registrations which are then reflected in different DNS name resolution hierarchies, then this uniqueness is lost. Instead, of everyone seeing the same site when they typed www.kafka.law into their browser, results will vary. What results a user will get will ordinarily depend on someone's choice, but that someone may be the user, or someone upstream from the user, depending on who selects the DNS. So far, however, the main consequence of alternate roots is that they create a need either for user education, or for the DNS equivalent of area codes. In a world of thriving competitive roots with name collisions, however, even users who controls their DNS service sometimes might experience unexpected results if, for example, they were to use a web-based e-mail form to send mail. Mail to fred@kafka.law on fred's machine might go somewhere different than email to that address sent from somewhere else. Indeed, to the extent that Internet services rely on intermediate machines for their transport via domain names rather than IP numbers, the routing of the data may vulnerable to any routing errors induced by inconsistent DNS resolution en route.

---

[212]  For an example of an alternate root operator who runs colliding TLDs, see  Sarah Ferguson, *Casting a Wider Net*, Village Voice (Apr. 10, 2001) (profiling Paul Garrin of Name.space), http://www.villagevoice.com/issues/0114/ferguson.php

[213]  Until now there have been no alternate roots open to the public that carry data conflicting with the legacy root.  Thus, when a user of an alternate root types a name in .com, .edu or .uk, that user gets the same IP number as does a user of the legacy root.  However, the recent introduction of a .biz TLD may change that.  Many of the existing alternate roots use a root file that points to a small .biz registry operated by Atlantic Root Network, http://www.biztld.net/.  If they persist in doing so after the ICANN-sponsored .biz goes live, the supersets will become conflicting sets.

[214]  *See* Mark A. Lemley & David McGowan, *Legal Implications of Network Economic Effects*, 86 **Calif. L. Rev.** 479, 553 (1998).

[215]   *See supra* note 94 (Cerf comments); cf. Brian Carpenter, IAB Technical Comment on the Unique DNS Root, http://www.iab.org/iab/IAB-Technical-Comment.txt.

[216]  Usually, but not always, a resource at a single IP number.

None of these dangers occur, however, in the absence of TLD name collisions. A merely supplementary alternate root does not present the same dangers, so long as there are no name collisions within it either. Even here, however, users may be frustrated if, much like a person trying to reach a new.net TLD from the legacy root today, they click on a link and get an error message because their DNS does not recognize the existence of the supplementary TLD.

An additional, potentially more serious, problem is 'cache poisoning', which can occur without TLD name collisions, and indeed without alternate roots. The DNS uses a number of shortcuts to allow DNS servers to cache data and quickly resolve domain names to IP numbers. So-called cache poisoning occurs when some of the extra data sent along in response to a DNS resolution query points somewhere other than where the operator of the DNS resolver expects. ICANN argued that the increase in the use of alternate roots might worsen the problem.[217]

The significance of these various technical factors remains controversial. In May, 2000, the Internet Architecture Board, which functions as the IETF's steering committee weighed in against alternate roots, stating in RFC 2826,

> To remain a global network, the Internet requires the existence of a globally unique public name space. The DNS name space is a hierarchical name space derived from a single, globally unique root. This is a technical constraint inherent in the design of the DNS. Therefore it is not technically feasible for there to be more than one root in the public DNS. That one root must be supported by a set of coordinated root servers administered by a unique naming authority.
>
> Put simply, deploying multiple public DNS roots would raise a very strong possibility that users of different ISPs who click on the same link on a web page could end up at different destinations, against the will of the web page designers.[218]

M. Stuart Lynn's paper for ICANN echoed this language.[219]

---

[217] *See* M. Stuart Lynn, "ICP-3" http://www.icann.org/icp/icp-3.htm

[218] IAB, RFC 2826, http://www.ietf.org/rfc/rfc2826.txt

[219] *See* Lynn, *supra* note 217.

Ordinarily, a paper from the IAB would be considered all but authoritative. Nevertheless, supporters of alternate roots attacked RFC 2826 as political, and noted that as it was labeled "Informational" it had not been subjected to the IETF's consensus building processes. Subsequently, others proposed ways of organizing alternate roots with a degree of coordination that, they argue, would not create the problems which worry the IAB,[220] although their proposals have not been adopted by the IETF, and indeed the IETF has apparently refused to allow them to proceed to discussion. Most recently, ICANN's own Protocol Supporting Organization (PSO) was asked to opine on RFC 2826's condemnation of alternate roots. It refused to endorse it, instead issuing the an artful (draft) statement:

> The Internet DNS currently operates using a Single Authoritative Root Server System. Although, it would be technically possible to devise and standardize a fully compliant alternative multiple root server system, there appears no technical reason for changing from the present working system, as this would require the development of a new set of protocols for use by the DNS.[221]

The PSO's statement is artful because the "new set of protocols" to which it refers might mean any one of three things. In theory, the new protocols might be primarily *social* rather than *technical*: ICANN might find a way to co-exist with alternate roots and agree to avoid name collisions. Or, radically decentralizing technical protocols might be created that allowed users, or their software, to select among multiple roots much like people dial area codes to select among otherwise identical telephone numbers.[222] Or, ICANN could adopt new protocols that added new capabilities to the ICANN's existing hierarchical root. These new capabilities would allow new functions akin to alternate roots, although they would require new user software and would leave the DNS firmly in ICANN's control.[223]

ICANN's exclusionary conduct towards competitors is exemplified in its treatment of new.net. New.net is both more and less than a true alternate root, and today it is perhaps the most visible competitor for namespace with ICANN.[224] New.net markets itself as a source of domains in thirty new English-language TLDs with names such as .shop, .kids, .law, .xxx, plus a large number of attractive Spanish, French and Portuguese TLDs.[225]

---

[220] *E.g.* S. Higgs, Alternative Roots and the Virtual Inclusive Root (May, 2001), http://www.ietf.org/internet-drafts/draft-higgs-virtual-root-00.txt; S. Higgs, Root Server Definitions (May, 2001), http://www.ietf.org/internet-drafts/draft-higgs-root-defs-01.txt; Karl Auerbach, Delving Into Multiple DNS Roots (MS word file, undated) http://www.cavebear.com/tmp/multiple-roots.doc

[221] http://www.pso.icann.org/PSO_Minutes/PSO-Minutes-4Sep2001.txt

[222] *See supra* note 220.

[223] For example, M. Stuart Lynn's proposal for experimental alternate roots relies on the use of the creation of new 'class identity' identifiers, and software to resolve the new class(es). The structure of the DNS, and the hierarchical control over the root, would remain exactly as it is today. *See* M. Stuart Lynn, ICP 3, *supra* note 217.

[224] On the growth of new.net, see May Wong, *Rebel Registry Adds 20 Domain Name Extensions*, **S.F. Chron.**, March 6, 2001, at C3; Chris Gaither, *New Challenge to Domain Name Registry*, **N.Y. Times**, May 15, 2001, at C10 (noting that Prodigy now supports new.net).

[225] *See* http://www.new.net

The registrant of, say, kafka.law at new.net actually receives a dual registration. In addition to receiving kafka.law in the New.net DNS, she also receives a registration of kafka.law.new.net in the legacy root -- a fourth-level sub-domain of new.net. Since New.net's .law domain is not in the root, today most Internet users worldwide who attempt to access kafka.law will get an error message. New.net attempts to overcome this, and simulate a genuine legacy TLD, by using a combination of two strategies, one aimed at Internet Service Providers (ISPs) and one aimed at users. New.net invites (and perhaps even pays) ISPs to alter their DNS to include new.net's TLDs.[226] To date new.net claims it has agreements with ISPs with more than 60 million users, a significant number, but only a fraction of the more than 400 million estimated Internet users worldwide.[227] For everyone else, new.net offers a 'plug-in' program that users of popular browsers can install on their computers. Once this program is installed it intercepts attempts to access any new.net TLD (or to email to a new.net address) and adds the "new.net" extension as needed. Thus, users of the plug-in and customers of participating ISPs can browse the both the legacy namespace and the new.net namespace at will. For them, www.kafka.law will resolve, and mail to kafka.law will reach its destination (albeit as kafka.law.new.net in some cases). Difficulties start, however, when the holder of the kafka.law registration wants to have a person who neither has a participating ISP nor the plug-in write back or visit her new web site. They must either type the full legacy address of kafka.law.new.net -- which more or less defeats the purpose of having the catchy name in the first place -- or be induced to get the plug-in.[228] And if they use a browser or operating for which there is no plug-in, even that is not an option. Obviously, new.net is hoping to break through the network effect and get more people to become part of its network. Equally obviously, it has yet to work; as one wag put it 'you don't see those TLDs on business cards.'

Despite new.net's limited market penetration, ICANN has singled it out for vituperative criticism and crafted new policies designed to ensure that potential customers understand their new.net registrations will never be recognized in the ICANN root. First, ICANN's Chief Policy Officer accused new.net of "breaking the Internet" and "selling snake oil".[229] Then, ICANN's CEO authored a paper attacking new.net's bona fides and legitimacy that he (eventually[230]) labeled a "discussion draft".[231] Then, without any warning, ICANN announced that a slightly revised version of the paper was official ICANN policy, and that no 'bottom-up' discussions were required because the paper merely 'restated' long-standing policy rather than making it.[232] In fact, however, the paper contained a number of new policies designed to make clear to the Internet community that ICANN had no intention of allowing new.net domains into the root, and indeed would feel very free to create colliding TLDs if and when it chose. These conclusions were both novel and controversial, and a reconsideration request to ICANN is currently pending.[233]

---

[226] *See* New.net, Information for ISPs, http://www.new.net/help_isp_info.tp; Aaron Hopkins, Re: new.net: yet another DNS namespace overlay play, http://www.merit.edu/mail.archives/html/nanog/2001-03/msg00136.html (Mar 7, 2001).

[227] *See How Many Online?* http://www.nua.ie/surveys/how_many_online/index.html ('educated guess' as of Dec., 2000).

[228] *See* New.net, FAQ, "Can I use my New.net domain names for e-mail?", http://www.new.net/help_faq.tp#e1.

[229] Kevin Murphy, ICANN Strikes Back, Refuses to be Strong-Armed by New.net, Network Briefing Daily: Issue , July 12, 2001, http://www.softwareuncovered.com/news/nbd-20010712.html (comments of ICANN Chief Policy Officer Andrew McLaughlin); http://www.icann.org/correspondence/schecter-letter-to-icann-16jul01.htm.

[230] There was some initial confusion on this as ICANN initially published the paper on its web page without any sign that it was a draft, a personal statement, or for discussion. After a brief storm of protest, ICANN added a preface from ICANN CEO M. Stuart Lynn saying it was his attempt "to restate existing policy and the technical basis for such policy." See ICANN's Lynn on alternative roots, ICANNWatch.org, http://www.icannwatch.org/article.php?sid=180.

[231] http://www.icann.org/stockholm/unique-root-draft.htm

ICANN needed a new policy because new.net presents a substantial potential threat to ICANN's monopoly over the TLD namespace. Were new.net to achieve critical mass in a TLD, ICANN's would find it difficult to create a colliding TLD without facing accusations that it, as the latecomer to that name, was the one 'breaking the Internet' by creating name conflicts for a substantial installed base of users. Worse from ICANN's point of view, new.net has grabbed some of the most popular TLDs (often after asking potential users to vote on which TLDs they'd like to see created). From ICANN's perspective as a self-described guardian of the public trust, it is wrong to allow an entrepreneur to grab whatever attractive names it wants rather than taking its chances along with other applicants to ICANN.[234] And indeed, new.net has chosen many TLDs that collide with longstanding 'true' alternate roots.

ICANN's reaction to New.net has been very different from its treatment of non-ASCII internationalized domain names, a development that poses at least as great a risk of non-unique domain names, at least from the users' perspective. ICANN has supported efforts to create non-ASCII domain names despite the market's adoption of competing encoding conventions. Currently these competing would-be standards amount to a series of parallel (i.e. alternate) domain name spaces. They differ from alternate roots in two ways. First, the internationalized domain name encoding schemes are usually labeled 'testbeds' and 'experiments'; ICANN takes the view that once a single standard emerges from the relevant standards

---

[232] *See* http://www.icann.org/icp/icp-3-background/lynn-statement-09jul01.htm; see also Jonathan Weinberg. How ICANN Policy Is Made, http://www.icannwatch.org/article.php?sid=241.

[233] *See* http://www.icannwatch.org/article.php?sid=286 (describing request from Jonathan Weinberg and Michael Froomkin).

[234] "Some of these operators and their supporters assert that their very presence in the marketplace gives them preferential right to TLDs to be authorized in the future by ICANN. They work under the philosophy that if they get there first with something that looks like a TLD and invite many registrants to participate, then ICANN will be required by their very presence and force of numbers to recognize in perpetuity these pseudo TLDs, inhibiting new TLDs with the same top-level name from being launched through the community's processes.

"No current policy would allow ICANN to grant such preferential rights. To do so would effectively yield ICANN's mandate to introduce new TLDs in an orderly manner in the public interest to those who would simply grab all the TLD names that seem to have any marketplace value, thus circumventing the community-based processes that ICANN is required to follow. For ICANN to yield its mandate would be a violation of the public trust under which ICANN was created and under which it must operate. Were it to grant such preferential rights, ICANN would abandon this public trust, rooted in the community, to those who only act for their own benefit. Indeed, granting preferential rights could jeopardize the stability of the DNS, violating ICANN's fundamental mandate."
M. Stuart Lynn, *supra* note 217.

body, all the competing encoding schemes should conform to the standard.[235] As these alternate encoding schemes are being widely deployed and sold, however, there is a real potential for the creation of de facto competing and colliding name spaces in non-romance character sets. A second difference is technical: non-ASCII coding schemes involve using some program to transliterate characters (e.g. Kanji) to a set of ASCII characters before sending a DNS request to resolve a domain name. Thus, the actual domain name registration remains an ASCII-character registration in the ICANN-controlled root. While the domain names may be 'alternate' and even 'conflicting' in the eye of the beholder, from ICANN's perspective they remain 'unique' and under its ultimate control.

It is not clear how courts would evaluate all of this in the context of an exclusive dealing antitrust claim. On the one hand, it is well-established that only procompetitive arguments may be considered as legitimate business justifications. ICANN is not free to argue that its foreclosure of competition was a good thing because competition itself is undesirable.[236] On the other hand, courts are willing to consider certain justifications for the regularization of competition in circumstances in which a market might not otherwise form. Thus, in *BMI v. CBS*,[237] the Court permitted a copyright owners cartel that provided licenses to millions of songs at a single flat rate. The Court reasoned that the cartel itself was procompetitive, since it eliminated transactions costs that would otherwise be prohibitive, and effectively "made" a new market. Similarly, some courts have permitted stock exchanges and trade associations to set up internal rules governing who can participate and excluding outsiders where such rules were necessary to let the market function effectively.[238] Such restrictions are not always permitted, however, and the courts will inquire in detail into whether the restriction on competition is actually necessary.[239]

It is not clear whether the desire for DNS uniformity justifies ICANN's exclusion of alternate roots from the list of potential registries. Certainly, the antitrust cases suggest that ICANN's asserted justifications will be subject to searching scrutiny on their merits.

---

[235] See ICANN, Board Resolution on Multi-Lingual Domain Names (Sep. 25, 2000), http://www.icann.org/minutes/minutes-25sep00.htm#MultilingualDomainNames; ICANN, Status Report of the Internationalized Domain Names Internal Working Group of the ICANN Board of Directors (June 1, 2001), http://www.icann.org/committees/idn/status-report-05jul01.htm.

[236] *See* National Society of Professional Engineers v. United States, 435 U.S. 679, 689-90 (1978).

[237] 441 U.S. 1 (1979).

[238] *See, e.g.,* Chicago Board of Trade v. United States, 246 U.S. 231 (1922). *Cf.* II **Herbert Hovenkamp et al., IP and Antitrust** §35.3 (forthcoming 2001) (discussing legitimate reasons to impose membership restrictions on standards bodies and related groups).

[239] *See, e.g.,* NCAA v. Regents of the University of Oklahoma, 468 U.S. 85 (1984) (rejecting NCAA's justification for limits on television coverage of college football); *United States v. Realty Multi-List, Inc.*, 629 F.2d 1351, 1369-1387 (5th Cir. 1980) (conducting an exhaustive inquiry into a real estate association's membership rules, and concluding that they violated the rule of reason because the organization had market power and the rules were insufficiently related to legitimate business concerns).. *Cf.* Silver v. New York Stock Exchange, 373 U.S. 341 (1963) (stock exchange violated antitrust laws by excluding members without providing them notice and a hearing).

### D. The Uniformity of the UDRP

Competition is not only about selling goods or services at the lowest price. For competition to be free and unfettered, companies must be able to compete as well on the nature and quality of the products they sell. Competition in the breakfast cereal industry, for example, requires not just that many different companies produce corn flakes, but that different companies be free to experiment with different types of cereal. Companies in that industry clearly establish market niches in part on their willingness to serve different types of customers with different types of cereals.

So too with goods or services of any type, including domain names. Registrars in a competitive marketplace will attempt to take business away from each other not only by lowering their price, but also by offering different and better services than their competition. Because the registrars' customers are domain name registrants, registrars in a competitive market might be expected to compete by offering rights or benefits that make their domain names more valuable. Among the things that registrars would compete over are the way, speed, and skill with which they would resolve domain name trademark disputes. Indeed, those registrars that predate ICANN had different policies for dealing with such disputes.

The UDRP short-circuited this competition. ICANN required all registrars to agree to impose a uniform dispute resolution policy on their registrants. By doing so, ICANN entered into a vertical agreement restricting non-price competition on one axis. This in and of itself is not necessarily an antitrust problem. Manufacturers regularly impose non-price restraints on their distributors or retailers; doing so may legitimately serve to prevent free riding and is normally legal.[240]

More troubling is the means by which the UDRP was adopted. ICANN did not develop the UDRP itself and impose it on the registrars. Rather, a group of registrars themselves banded together and drafted the initial provisions with input from intellectual property owners. These registrars then collectively presented the draft to ICANN, which adopted it with only minor changes. Thus, it appears that the UDRP was not in fact merely a vertical agreement imposed by ICANN on its customers, but actually reflects a horizontal agreement among the registrars themselves to limit competition in dispute resolution procedures. Horizontal agreements are much more worrisome, particularly where (as here) they are entered into by the largest companies in the market.[241] ICANN appears not to have been the driving force in drafting the policy, but rather a "ringmaster" employed by the registrars to enforce their own agreement.[242] The issue is more complicated, however, because the registrars in turn have no direct incentive to insist on uniform dispute resolution. Rather, they were motivated by pressure from the trademark owners, backed by the threat of lawsuits and a fear that the trademark constituency could prevent both registrar competition and the development of new gTLDs altogether. Michael Palage, the head of the Registrars' DNSO

---

[240] *See, e.g.,* Business Elec. v. Sharp Elec., 485 U.S. 717 (1988); Continental T.V., Inc. v. GTE Sylvania Inc., 433 U.S. 36 (1977).

[241] *See supra* note 81 (listing participants, including NSI).

[242] *See* Thomas G. Krattenmaker & Steven C. Salop, *Anticompetitive Exclusion: Raising Rivals' Costs to Achieve Power Over Price*, 96 **Yale L.J.** 209, 238-40, 260-62 (1986) (describing this as the "cartel ringmaster" theory).

Constituency, famously said, "The trademark lobby must be placated because of its potential ability and inclination to bankrupt new registrars and wreck havoc on their registrant databases."[243]

      Horizontal agreements to restrict non-price competition are not necessarily illegal per se. Rather, they will be given a "quick look" to determine whether there are legitimate procompetitive justifications for the agreement. Here, the obvious purpose of the agreement is to limit cybersquatting. There is strong evidence that the UDRP was enacted at the behest of intellectual property owners who likely had the political power to prevent the adoption of any new gTLDs unless the registrars agreed to restrict cybersquatters. Certainly the effect of the UDRP has been to punish cybersquatters, in part by establishing procedures that have systematically favored intellectual property owners even in doubtful cases.[244] But even granting that cybersquatting is a bad thing, collusion among erstwhile competitors to treat it uniformly is not necessarily legal. The Supreme Court has made it clear that justifications for horizontal agreements must be procompetitive, not just good social policy. It has rejected justifications for cartels based on the idea that competition itself will lead to bad results.[245] Here, the clear effect of the UDRP is to eliminate competition that otherwise would have existed between registrars about how to resolve disputes. That competition may well have been undesirable as a matter of social policy.[246] But as a matter of antitrust law, it doesn't matter. There does not seem to be the sort of market-making necessity for the UDRP that ICANN has asserted as a justification for excluding alternate roots.[247]

---

[243] *See* Judith Oppenheimer, Adventive, http://www.judithoppenheimer.com/pressetc/adentive.html (quoting remark by Palage at a January 10, 2000 Small Business Administration meeting on Domain Name Issues).

[244] For empirical evidence that this has happened, see Michael Geist, *Fair.com? An Examination of the Allegations of Systematic Unfairness in the ICANN UDRP* (working paper 2001) (http://aix1.uottawa.ca/~geist/geistudrp.pdf); Milton Mueller, *Rough Justice: An Analysis of ICANN's Uniform Dispute Resolution Policy* (working paper 2000) (http://dcc.syr.edu/roughjustice.htm).

[245] *See, e.g.,* National Society of Professional Engineers v. United States, 435 U.S. 679, 689-90 (1978) ("The early cases also foreclose the argument that because of the special characteristics of a particular industry, monopolistic arrangements will better promote trade and commerce than competition. That kind of argument is properly addressed to Congress and may justify an exemption from the statute for specific industries, but it is not permitted by the Rule of Reason.").

[246] There is a great deal of academic debate over whether certain forms of regulatory competition result in a "race to the bottom," in which regulatory regimes are rewarded for being inefficiently lax. *See, e.g.,* Lynn M. LoPucki & Sara D. Kalin, *The Failure of Public Company Bankruptcies in Delaware and New York: Empirical Evidence of a "Race to the Bottom"*, 54 **Vand. L. Rev.** 231, 237 (2001); **Roberta Romano, The Genius of American Corporate Law** (1993); Ehud Kamar, *A Regulatory Competition Theory of Indeterminacy in Corporate Law*, 98 **Colum. L. Rev.** 1908 (1998); Jonathan R. Macey & Geoffrey P. Miller, *Toward an Interest-Group Theory of Delaware Corporate Law*, 65 **Texas L. Rev.** 469 (1987). While registrars are not governments, one might reasonably fear a similar effect in a competitive regime, since registrants (registrars' customers) might prefer not to be subject to any sort of dispute resolution scheme at all, but rather to externalize any costs of their trademark infringement.

    The matter is complicated in this case by the fact that the U.S. Congress passed the Anticybersquatting Consumer Protection Act, 15 U.S.C. § 1125(d), the same month the UDRP was adopted. Thus, dispute resolution competition among registrars would have been limited in any event by the legal backstop; trademark owners were and remain free to go to court rather than use any private dispute resolution system.

[247] *See supra* notes __-__ and accompanying text.

ICANN's liability for adopting the UDRP is related to antitrust concerns about its policy on alternate roots as well. Alternate roots are not subject to the UDRP because they have not contracted with ICANN. They therefore constitute a potential source of competition in registration policies, one that ICANN is foreclosing. Thus, neither policy should be considered in isolation.

If the standardized UDRP agreement is illegal, ICANN is liable regardless of whether it was the motivating force behind the policy. Even reluctant or coerced co-conspirators violate the antitrust laws by entering into the conspiracy.[248] Further, the standard-setting cases seem to suggest that standard-setting organizations themselves violate the antitrust laws even if their processes are hijacked by interested participants.[249] But the more likely defendants in a UDRP claim are the registrars who collectively drafted and implemented it. We turn to their possible liability in the next section.

## IV.    Liability for Petitioning ICANN

In addition to the state action and governmental immunity doctrines, private actors who petition the government in the effort to influence it to act are immune from antitrust liability even if the actions they seek are anticompetitive. This "*Noerr-Pennington*" immunity[250] creates a sort of penumbra around the state action doctrine in which anticompetitive petitioning may take place without antitrust liability. The fundamental basis for this petitioning immunity is the First Amendment right to petition.[251] As Justice Scalia put it, it would be "peculiar in a democracy, and perhaps in derogation of the constitutional right 'to petition the government for a redress of grievances' . . . to establish a category of lawful state action that citizens are not permitted to urge."[252] Efforts to petition the government are immune from antitrust liability unless those efforts amount to no more than a "sham."[253]

In this case, though, registrars, registrants and trademark owners are not petitioning the government itself. Rather, they are petitioning ICANN, a private corporation acting under the authorization of the government. There is of course no immunity for "petitioning" a purely private entity; were it otherwise, cartels and other agreements between competitors would all be immune.[254] But petitioning ICANN is perhaps an intermediate case, since ICANN's existence is authorized by governmental policy, and there is a plausible argument that it is a state actor for antitrust purposes.[255] The closest analogue is *Allied Tube & Conduit Corp. v. Indian Head Inc.*,[256] which involved efforts to influence the National Fire Protection Association, a private standard-setting body whose model codes were routinely enacted into law unchanged by state legislatures.[257] The Court held that Allied Tube was not immune from liability for petitioning the NFPA because the NFPA lacked the public accountability of a truly governmental body:

> Whatever de facto authority the Association enjoys, no official authority has been conferred on it by any government, and the decisionmaking body of the Association is composed, at least in part, of persons with economic incentives to restrain trade. . . . [W]here, as here, the restraint is imposed by persons unaccountable to the public and without official authority, many of whom have personal financial interests in restraining competition, we have no difficulty concluding that the restraint has resulted from private action.[258]

---

[248] Calnetics Corp. v. Volkswagen of America, Inc., 532 F.2d 674, 682 (9th Cir.) ("The involuntary nature of one's participation in a conspiracy to monopolize is no defense."), cert. denied, 429 U.S. 940 (1976); cf. MCM Partners Inc. v. Andrews-Bartlett & Assocs., Inc., 62 F.3d 967, 973 (7th Cir. 1995) (citing the Supreme Court in United States v. Paramount Pictures, Inc., 334 U.S. 131, 161 (1948), and thirteen other cases, to support its holding that a Section 1 conspiracy "is not negated by the fact that one or more of the co-conspirators acted unwillingly, reluctantly, or only in response to coercion").

[249] *See, e.g.,* American Society of Mechanical Engineers v. Hydrolevel Corp., 456 U.S. 556 (1982) (ASME violated antitrust laws where a member sent a threatening letter to one of its competitors on ASME letterhead, even though the member acted without actual authority in sending the letter).

[250] *See, e.g.,* Eastern Railroad Presidents Conference v. Noerr Motor Freight, Inc., 365 U.S. 127 (1961); United Mine Workers v. Pennington, 381 U.S. 657 (1965).

[251] On the source of petitioning immunity, *see* McGowan & Lemley, *supra* note __, at 307-14.

[252] City of Columbia v. Omni Outdoor Advertising, 499 U.S. 365, 379 (1991).

[253] *See* Professional Real Estate Investors v. Columbia Pictures Inc., 508 U.S. 49 (1993).

To the extent that ICANN lacks public accountability,[259] has no official authority,[260] and has a financial interest to restrain competition,[261] it is a private body and those who petition it will enjoy no special immunity.[262]

As with the state action doctrine, merely determining that immunity does not apply is only the beginning of the inquiry. Courts must still apply the normal principles of antitrust law to determine whether the act of petitioning is itself anticompetitive. If the petitioners are the registrars seeking to have ICANN impose a uniform UDRP, that analysis was conducted above. Alternatively, one might suggest that certain trademark owners themselves were liable for petitioning ICANN to help them out by setting favorable dispute resolution rules. The problem with this theory is that trademark owners aren't really "competing" in any relevant market; they simply want the rules changed in a way that is more favorable to them.

---

[254] *See, e.g.,* Raymond Ku, *Antitrust Immunity, the First Amendment, and Settlements: Defining the Boundaries of the Right to Petition*, 33 **Ind. L. Rev.** 385 (2000).

[255] As noted above, however, we ultimately find that argument unpersuasive. *See supra* notes __-__ and accompanying text.

[256] 486 U.S. 492 (1988).

[257] *Id*. at 495.

[258] *Id*. at 501-02.

[259] Many arguments have been made along these lines. *See, e.g.,* Froomkin, *supra* note 5; Liu, *supra* note 7.

[260] On this more complicated issue, see *supra* notes __-__ and accompanying text; *see generally* Froomkin, *supra* note 5.

[261] *See supra* notes __-__ and accompanying text (describing ICANN's benefit from restraining competition by alternate roots). ICANN seems to lack a similar financial incentive to enforce the UDRP, except to the extent its adoption placates trademark owners who would otherwise have the power to block ICANN initiatives.

[262] *See also* Sessions Tank Liners v. Joor Mfg. Co., 17 F.3d 295 (9th Cir. 1994) (misrepresentations to a quasi-public standard-setting organization could violate the antitrust laws where the organization acted in an administrative rather than a legislative capacity). For a similar argument relating to NSI's predecessor, the InterNIC, see Stephen J. Davidson & Nicole A. English, *Applying The Trademark Misuse Doctrine To Domain Name Disputes* (1996), http://www.cla.org/t_misuse.htm.

More generally, if private actors do not enjoy petitioning immunity for their contacts with ICANN, they will have to take more care than they have to date to conform their behavior to the requirements of antitrust law. Groups of registrars, registries, or potential registries must take particular care about agreeing together on a course of conduct; section 1 of the Sherman Act imposes greater restrictions on horizontal agreements to restrict trade than the restrictions on unilateral conduct we have discussed so far.

## V. Policy Implications

By delegating policy-making authority to ICANN, a private actor, without putting in place any real mechanisms for accountability, the government has created some unanticipated legal problems. It seems clear that the government itself could operate the legacy root in a way that excludes alternate roots without violating the antitrust laws. Similarly, the government should be able to impose a uniform domain name dispute resolution policy on registrars and registrants without antitrust liability.[263] It could also delegate these tasks to a private entity like ICANN without antitrust liability if the government affirmatively set the policy and actively supervised ICANN's implementation of it. Under *Noerr* and the government immunity doctrines, the price of unsupervised delegation is antitrust scrutiny. And it is not clear that ICANN and those in a position to influence it will survive that scrutiny. To avoid antitrust liability, ICANN will have to consider carefully both its policy regarding alternate roots and likely its uniform dispute resolution policy as well. These policies are not necessarily illegal, but ICANN will have to offer evidence that they are on balance good for competition – something that to date it has not been obliged to do. At a minimum, ICANN's policies will be subject to increased scrutiny, and likely to protracted antitrust litigation.

This may not be a desirable result as a policy matter. There are plausible reasons to concentrate control of the root in one entity. Decentralized roots increase the chance of collisions or incompatibilities between TLDs operated by different entities. While there may be decentralized solutions to this problem, it is a risk that we might decide is not worth taking. Similarly, the UDRP performs a function many value: it gives trademark owners and domain name registrants a cheap and quick way to resolve disputes over alleged cybersquatting. The astonishing number of UDRP proceedings to date – over 4,000 – is a testament both to the continuing seriousness of the issue and the relative cost and speed of the UDRP compared to judicial action.[264] The UDRP has problems – it may not give respondents enough process or gather enough information, and there is recent evidence that it is systematically biased in favor of trademark owners.[265] But we might decide that a cheap and rapid dispute resolution system is worth giving up some certainty that the outcomes are correct.

---

[263] As one of us has noted elsewhere, however, the nature of the UDRP might raise constitutional concerns if compelled directly by a state actor. *See* Froomkin, *supra* note 5, at 96-101, 136-38.

[264] A search for "Anticybersquatting Consumer Protection Act" on Westlaw in September 2001 turned up 60 cases. By contrast, in the same period UDRP arbitrators decided 3,988 cases involving 7,009 different domain names. Hundreds more were pending.

The problem, though, is that as it stands presently "we" don't get to make any such decision. Whether to allow alternate roots, and how to design a domain name dispute resolution policy, are important policy questions. They may be decisions we ought to leave to the market, an approach that would allow alternate roots and would permit registrars to design non-uniform dispute resolution policies. Alternatively, the government may decide that it should displace the market outcome in the interest of ensuring the stability of the Internet or dealing with the problem of cybersquatting.[266] In either case, the decision will have been made by an institution that is accountable to the public in some form – either the market, in which consumers can "vote with their wallet," or the government. Accountability is desirable because it permits error correction. If it turns out that the UDRP system has structural flaws, for example – as appears to be the case[267] – those flaws can be identified and corrected. By contrast, government today abdicates this decision to a private, unelected entity that is also not subject to normal market constraints. If it turns out that ICANN makes the wrong decision, there is currently nothing to be done.

The government should take a more active role in setting domain name policy, either by running the DNS itself, or by actively supervising its delegates, or by making an affirmative decision to let the market work unfettered. If the government won't step in to do one of these things, ICANN and those who deal with it will find many of their current activities constrained by antitrust rules.

---

[265] *See supra* note __ (citing such evidence).

[266] For an argument that the questions ICANN is addressing are policy questions that are public in nature, see Liu, *supra* note 7, at 604. For even stronger suggestions that privatizing the network may be inefficient, see Brett Frischmann, *Rethinking Privatization and Commercialization of the Internet* (working paper 2001); Kesan & Shah, *supra* note 5.

[267] *See, e.g.,* Geist, *supra* note 151; Mueller, *supra* note 151. Among the more obvious structural problems are the short time frame for response, which prevents many respondents from answering at all or from retaining a lawyer; the lack of an appeal procedure; and the fact that complainants get to select the private company that will arbitrate the dispute, giving those companies every incentive to cater to complainants (trademark owners) in deciding cases. For more detail on these procedural deficiencies, see Froomkin, *supra* note 5, at 96-101; Froomkin, *supra* note 87.

51